\documentclass[useAMS,usenatbib,referee]{mn2e}
\usepackage{times}
\usepackage{url}
\usepackage{graphicx}
\usepackage{deluxetable}
\usepackage[usenames]{color}
\DeclareGraphicsExtensions{.pdf,.png,.jpg,.mps,.eps,.ps}
\usepackage{amsmath}
\usepackage{natbib}
\usepackage{subfig}
\usepackage{soul}
\definecolor{AliceBlue}{rgb}{0.94,0.97,1.00}
\definecolor{AntiqueWhite1}{rgb}{1.00,0.94,0.86}
\definecolor{AntiqueWhite2}{rgb}{0.93,0.87,0.80}
\definecolor{AntiqueWhite3}{rgb}{0.80,0.75,0.69}
\definecolor{AntiqueWhite4}{rgb}{0.55,0.51,0.47}
\definecolor{AntiqueWhite}{rgb}{0.98,0.92,0.84}
\definecolor{BlanchedAlmond}{rgb}{1.00,0.92,0.80}
\definecolor{BlueViolet}{rgb}{0.54,0.17,0.89}
\definecolor{CadetBlue1}{rgb}{0.60,0.96,1.00}
\definecolor{CadetBlue2}{rgb}{0.56,0.90,0.93}
\definecolor{CadetBlue3}{rgb}{0.48,0.77,0.80}
\definecolor{CadetBlue4}{rgb}{0.33,0.53,0.55}
\definecolor{CadetBlue}{rgb}{0.37,0.62,0.63}
\definecolor{CornflowerBlue}{rgb}{0.39,0.58,0.93}
\definecolor{DarkBlue}{rgb}{0.00,0.00,0.55}
\definecolor{DarkCyan}{rgb}{0.00,0.55,0.55}
\definecolor{DarkGoldenrod1}{rgb}{1.00,0.73,0.06}
\definecolor{DarkGoldenrod2}{rgb}{0.93,0.68,0.05}
\definecolor{DarkGoldenrod3}{rgb}{0.80,0.58,0.05}
\definecolor{DarkGoldenrod4}{rgb}{0.55,0.40,0.03}
\definecolor{DarkGoldenrod}{rgb}{0.72,0.53,0.04}
\definecolor{DarkGray}{rgb}{0.66,0.66,0.66}
\definecolor{DarkGreen}{rgb}{0.00,0.39,0.00}
\definecolor{DarkGrey}{rgb}{0.66,0.66,0.66}
\definecolor{DarkKhaki}{rgb}{0.74,0.72,0.42}
\definecolor{DarkMagenta}{rgb}{0.55,0.00,0.55}
\definecolor{DarkOliveGreen1}{rgb}{0.79,1.00,0.44}
\definecolor{DarkOliveGreen2}{rgb}{0.74,0.93,0.41}
\definecolor{DarkOliveGreen3}{rgb}{0.64,0.80,0.35}
\definecolor{DarkOliveGreen4}{rgb}{0.43,0.55,0.24}
\definecolor{DarkOliveGreen}{rgb}{0.33,0.42,0.18}
\definecolor{DarkOrange1}{rgb}{1.00,0.50,0.00}
\definecolor{DarkOrange2}{rgb}{0.93,0.46,0.00}
\definecolor{DarkOrange3}{rgb}{0.80,0.40,0.00}
\definecolor{DarkOrange4}{rgb}{0.55,0.27,0.00}
\definecolor{DarkOrange}{rgb}{1.00,0.55,0.00}
\definecolor{DarkOrchid1}{rgb}{0.75,0.24,1.00}
\definecolor{DarkOrchid2}{rgb}{0.70,0.23,0.93}
\definecolor{DarkOrchid3}{rgb}{0.60,0.20,0.80}
\definecolor{DarkOrchid4}{rgb}{0.41,0.13,0.55}
\definecolor{DarkOrchid}{rgb}{0.60,0.20,0.80}
\definecolor{DarkRed}{rgb}{0.55,0.00,0.00}
\definecolor{DarkSalmon}{rgb}{0.91,0.59,0.48}
\definecolor{DarkSeaGreen1}{rgb}{0.76,1.00,0.76}
\definecolor{DarkSeaGreen2}{rgb}{0.71,0.93,0.71}
\definecolor{DarkSeaGreen3}{rgb}{0.61,0.80,0.61}
\definecolor{DarkSeaGreen4}{rgb}{0.41,0.55,0.41}
\definecolor{DarkSeaGreen}{rgb}{0.56,0.74,0.56}
\definecolor{DarkSlateBlue}{rgb}{0.28,0.24,0.55}
\definecolor{DarkSlateGray1}{rgb}{0.59,1.00,1.00}
\definecolor{DarkSlateGray2}{rgb}{0.55,0.93,0.93}
\definecolor{DarkSlateGray3}{rgb}{0.47,0.80,0.80}
\definecolor{DarkSlateGray4}{rgb}{0.32,0.55,0.55}
\definecolor{DarkSlateGray}{rgb}{0.18,0.31,0.31}
\definecolor{DarkSlateGrey}{rgb}{0.18,0.31,0.31}
\definecolor{DarkTurquoise}{rgb}{0.00,0.81,0.82}
\definecolor{DarkViolet}{rgb}{0.58,0.00,0.83}
\definecolor{DeepPink1}{rgb}{1.00,0.08,0.58}
\definecolor{DeepPink2}{rgb}{0.93,0.07,0.54}
\definecolor{DeepPink3}{rgb}{0.80,0.06,0.46}
\definecolor{DeepPink4}{rgb}{0.55,0.04,0.31}
\definecolor{DeepPink}{rgb}{1.00,0.08,0.58}
\definecolor{DeepSkyBlue1}{rgb}{0.00,0.75,1.00}
\definecolor{DeepSkyBlue2}{rgb}{0.00,0.70,0.93}
\definecolor{DeepSkyBlue3}{rgb}{0.00,0.60,0.80}
\definecolor{DeepSkyBlue4}{rgb}{0.00,0.41,0.55}
\definecolor{DeepSkyBlue}{rgb}{0.00,0.75,1.00}
\definecolor{DimGray}{rgb}{0.41,0.41,0.41}
\definecolor{DimGrey}{rgb}{0.41,0.41,0.41}
\definecolor{DodgerBlue1}{rgb}{0.12,0.56,1.00}
\definecolor{DodgerBlue2}{rgb}{0.11,0.53,0.93}
\definecolor{DodgerBlue3}{rgb}{0.09,0.45,0.80}
\definecolor{DodgerBlue4}{rgb}{0.06,0.31,0.55}
\definecolor{DodgerBlue}{rgb}{0.12,0.56,1.00}
\definecolor{FloralWhite}{rgb}{1.00,0.98,0.94}
\definecolor{ForestGreen}{rgb}{0.13,0.55,0.13}
\definecolor{GhostWhite}{rgb}{0.97,0.97,1.00}
\definecolor{GreenYellow}{rgb}{0.68,1.00,0.18}
\definecolor{HotPink1}{rgb}{1.00,0.43,0.71}
\definecolor{HotPink2}{rgb}{0.93,0.42,0.65}
\definecolor{HotPink3}{rgb}{0.80,0.38,0.56}
\definecolor{HotPink4}{rgb}{0.55,0.23,0.38}
\definecolor{HotPink}{rgb}{1.00,0.41,0.71}
\definecolor{IndianRed1}{rgb}{1.00,0.42,0.42}
\definecolor{IndianRed2}{rgb}{0.93,0.39,0.39}
\definecolor{IndianRed3}{rgb}{0.80,0.33,0.33}
\definecolor{IndianRed4}{rgb}{0.55,0.23,0.23}
\definecolor{IndianRed}{rgb}{0.80,0.36,0.36}
\definecolor{LavenderBlush1}{rgb}{1.00,0.94,0.96}
\definecolor{LavenderBlush2}{rgb}{0.93,0.88,0.90}
\definecolor{LavenderBlush3}{rgb}{0.80,0.76,0.77}
\definecolor{LavenderBlush4}{rgb}{0.55,0.51,0.53}
\definecolor{LavenderBlush}{rgb}{1.00,0.94,0.96}
\definecolor{LawnGreen}{rgb}{0.49,0.99,0.00}
\definecolor{LemonChiffon1}{rgb}{1.00,0.98,0.80}
\definecolor{LemonChiffon2}{rgb}{0.93,0.91,0.75}
\definecolor{LemonChiffon3}{rgb}{0.80,0.79,0.65}
\definecolor{LemonChiffon4}{rgb}{0.55,0.54,0.44}
\definecolor{LemonChiffon}{rgb}{1.00,0.98,0.80}
\definecolor{LightBlue1}{rgb}{0.75,0.94,1.00}
\definecolor{LightBlue2}{rgb}{0.70,0.87,0.93}
\definecolor{LightBlue3}{rgb}{0.60,0.75,0.80}
\definecolor{LightBlue4}{rgb}{0.41,0.51,0.55}
\definecolor{LightBlue}{rgb}{0.68,0.85,0.90}
\definecolor{LightCoral}{rgb}{0.94,0.50,0.50}
\definecolor{LightCyan1}{rgb}{0.88,1.00,1.00}
\definecolor{LightCyan2}{rgb}{0.82,0.93,0.93}
\definecolor{LightCyan3}{rgb}{0.71,0.80,0.80}
\definecolor{LightCyan4}{rgb}{0.48,0.55,0.55}
\definecolor{LightCyan}{rgb}{0.88,1.00,1.00}
\definecolor{LightGoldenrod1}{rgb}{1.00,0.93,0.55}
\definecolor{LightGoldenrod2}{rgb}{0.93,0.86,0.51}
\definecolor{LightGoldenrod3}{rgb}{0.80,0.75,0.44}
\definecolor{LightGoldenrod4}{rgb}{0.55,0.51,0.30}
\definecolor{LightGoldenrodYellow}{rgb}{0.98,0.98,0.82}
\definecolor{LightGoldenrod}{rgb}{0.93,0.87,0.51}
\definecolor{LightGray}{rgb}{0.83,0.83,0.83}
\definecolor{LightGreen}{rgb}{0.56,0.93,0.56}
\definecolor{LightGrey}{rgb}{0.83,0.83,0.83}
\definecolor{LightPink1}{rgb}{1.00,0.68,0.73}
\definecolor{LightPink2}{rgb}{0.93,0.64,0.68}
\definecolor{LightPink3}{rgb}{0.80,0.55,0.58}
\definecolor{LightPink4}{rgb}{0.55,0.37,0.40}
\definecolor{LightPink}{rgb}{1.00,0.71,0.76}
\definecolor{LightSalmon1}{rgb}{1.00,0.63,0.48}
\definecolor{LightSalmon2}{rgb}{0.93,0.58,0.45}
\definecolor{LightSalmon3}{rgb}{0.80,0.51,0.38}
\definecolor{LightSalmon4}{rgb}{0.55,0.34,0.26}
\definecolor{LightSalmon}{rgb}{1.00,0.63,0.48}
\definecolor{LightSeaGreen}{rgb}{0.13,0.70,0.67}
\definecolor{LightSkyBlue1}{rgb}{0.69,0.89,1.00}
\definecolor{LightSkyBlue2}{rgb}{0.64,0.83,0.93}
\definecolor{LightSkyBlue3}{rgb}{0.55,0.71,0.80}
\definecolor{LightSkyBlue4}{rgb}{0.38,0.48,0.55}
\definecolor{LightSkyBlue}{rgb}{0.53,0.81,0.98}
\definecolor{LightSlateBlue}{rgb}{0.52,0.44,1.00}
\definecolor{LightSlateGray}{rgb}{0.47,0.53,0.60}
\definecolor{LightSlateGrey}{rgb}{0.47,0.53,0.60}
\definecolor{LightSteelBlue1}{rgb}{0.79,0.88,1.00}
\definecolor{LightSteelBlue2}{rgb}{0.74,0.82,0.93}
\definecolor{LightSteelBlue3}{rgb}{0.64,0.71,0.80}
\definecolor{LightSteelBlue4}{rgb}{0.43,0.48,0.55}
\definecolor{LightSteelBlue}{rgb}{0.69,0.77,0.87}
\definecolor{LightYellow1}{rgb}{1.00,1.00,0.88}
\definecolor{LightYellow2}{rgb}{0.93,0.93,0.82}
\definecolor{LightYellow3}{rgb}{0.80,0.80,0.71}
\definecolor{LightYellow4}{rgb}{0.55,0.55,0.48}
\definecolor{LightYellow}{rgb}{1.00,1.00,0.88}
\definecolor{LimeGreen}{rgb}{0.20,0.80,0.20}
\definecolor{MediumAquamarine}{rgb}{0.40,0.80,0.67}
\definecolor{MediumBlue}{rgb}{0.00,0.00,0.80}
\definecolor{MediumOrchid1}{rgb}{0.88,0.40,1.00}
\definecolor{MediumOrchid2}{rgb}{0.82,0.37,0.93}
\definecolor{MediumOrchid3}{rgb}{0.71,0.32,0.80}
\definecolor{MediumOrchid4}{rgb}{0.48,0.22,0.55}
\definecolor{MediumOrchid}{rgb}{0.73,0.33,0.83}
\definecolor{MediumPurple1}{rgb}{0.67,0.51,1.00}
\definecolor{MediumPurple2}{rgb}{0.62,0.47,0.93}
\definecolor{MediumPurple3}{rgb}{0.54,0.41,0.80}
\definecolor{MediumPurple4}{rgb}{0.36,0.28,0.55}
\definecolor{MediumPurple}{rgb}{0.58,0.44,0.86}
\definecolor{MediumSeaGreen}{rgb}{0.24,0.70,0.44}
\definecolor{MediumSlateBlue}{rgb}{0.48,0.41,0.93}
\definecolor{MediumSpringGreen}{rgb}{0.00,0.98,0.60}
\definecolor{MediumTurquoise}{rgb}{0.28,0.82,0.80}
\definecolor{MediumVioletRed}{rgb}{0.78,0.08,0.52}
\definecolor{MidnightBlue}{rgb}{0.10,0.10,0.44}
\definecolor{MintCream}{rgb}{0.96,1.00,0.98}
\definecolor{MistyRose1}{rgb}{1.00,0.89,0.88}
\definecolor{MistyRose2}{rgb}{0.93,0.84,0.82}
\definecolor{MistyRose3}{rgb}{0.80,0.72,0.71}
\definecolor{MistyRose4}{rgb}{0.55,0.49,0.48}
\definecolor{MistyRose}{rgb}{1.00,0.89,0.88}
\definecolor{NavajoWhite1}{rgb}{1.00,0.87,0.68}
\definecolor{NavajoWhite2}{rgb}{0.93,0.81,0.63}
\definecolor{NavajoWhite3}{rgb}{0.80,0.70,0.55}
\definecolor{NavajoWhite4}{rgb}{0.55,0.47,0.37}
\definecolor{NavajoWhite}{rgb}{1.00,0.87,0.68}
\definecolor{NavyBlue}{rgb}{0.00,0.00,0.50}
\definecolor{OldLace}{rgb}{0.99,0.96,0.90}
\definecolor{OliveDrab1}{rgb}{0.75,1.00,0.24}
\definecolor{OliveDrab2}{rgb}{0.70,0.93,0.23}
\definecolor{OliveDrab3}{rgb}{0.60,0.80,0.20}
\definecolor{OliveDrab4}{rgb}{0.41,0.55,0.13}
\definecolor{OliveDrab}{rgb}{0.42,0.56,0.14}
\definecolor{OrangeRed1}{rgb}{1.00,0.27,0.00}
\definecolor{OrangeRed2}{rgb}{0.93,0.25,0.00}
\definecolor{OrangeRed3}{rgb}{0.80,0.22,0.00}
\definecolor{OrangeRed4}{rgb}{0.55,0.15,0.00}
\definecolor{OrangeRed}{rgb}{1.00,0.27,0.00}
\definecolor{PaleGoldenrod}{rgb}{0.93,0.91,0.67}
\definecolor{PaleGreen1}{rgb}{0.60,1.00,0.60}
\definecolor{PaleGreen2}{rgb}{0.56,0.93,0.56}
\definecolor{PaleGreen3}{rgb}{0.49,0.80,0.49}
\definecolor{PaleGreen4}{rgb}{0.33,0.55,0.33}
\definecolor{PaleGreen}{rgb}{0.60,0.98,0.60}
\definecolor{PaleTurquoise1}{rgb}{0.73,1.00,1.00}
\definecolor{PaleTurquoise2}{rgb}{0.68,0.93,0.93}
\definecolor{PaleTurquoise3}{rgb}{0.59,0.80,0.80}
\definecolor{PaleTurquoise4}{rgb}{0.40,0.55,0.55}
\definecolor{PaleTurquoise}{rgb}{0.69,0.93,0.93}
\definecolor{PaleVioletRed1}{rgb}{1.00,0.51,0.67}
\definecolor{PaleVioletRed2}{rgb}{0.93,0.47,0.62}
\definecolor{PaleVioletRed3}{rgb}{0.80,0.41,0.54}
\definecolor{PaleVioletRed4}{rgb}{0.55,0.28,0.36}
\definecolor{PaleVioletRed}{rgb}{0.86,0.44,0.58}
\definecolor{PapayaWhip}{rgb}{1.00,0.94,0.84}
\definecolor{PeachPuff1}{rgb}{1.00,0.85,0.73}
\definecolor{PeachPuff2}{rgb}{0.93,0.80,0.68}
\definecolor{PeachPuff3}{rgb}{0.80,0.69,0.58}
\definecolor{PeachPuff4}{rgb}{0.55,0.47,0.40}
\definecolor{PeachPuff}{rgb}{1.00,0.85,0.73}
\definecolor{PowderBlue}{rgb}{0.69,0.88,0.90}
\definecolor{RosyBrown1}{rgb}{1.00,0.76,0.76}
\definecolor{RosyBrown2}{rgb}{0.93,0.71,0.71}
\definecolor{RosyBrown3}{rgb}{0.80,0.61,0.61}
\definecolor{RosyBrown4}{rgb}{0.55,0.41,0.41}
\definecolor{RosyBrown}{rgb}{0.74,0.56,0.56}
\definecolor{RoyalBlue1}{rgb}{0.28,0.46,1.00}
\definecolor{RoyalBlue2}{rgb}{0.26,0.43,0.93}
\definecolor{RoyalBlue3}{rgb}{0.23,0.37,0.80}
\definecolor{RoyalBlue4}{rgb}{0.15,0.25,0.55}
\definecolor{RoyalBlue}{rgb}{0.25,0.41,0.88}
\definecolor{SaddleBrown}{rgb}{0.55,0.27,0.07}
\definecolor{SandyBrown}{rgb}{0.96,0.64,0.38}
\definecolor{SeaGreen1}{rgb}{0.33,1.00,0.62}
\definecolor{SeaGreen2}{rgb}{0.31,0.93,0.58}
\definecolor{SeaGreen3}{rgb}{0.26,0.80,0.50}
\definecolor{SeaGreen4}{rgb}{0.18,0.55,0.34}
\definecolor{SeaGreen}{rgb}{0.18,0.55,0.34}
\definecolor{SkyBlue1}{rgb}{0.53,0.81,1.00}
\definecolor{SkyBlue2}{rgb}{0.49,0.75,0.93}
\definecolor{SkyBlue3}{rgb}{0.42,0.65,0.80}
\definecolor{SkyBlue4}{rgb}{0.29,0.44,0.55}
\definecolor{SkyBlue}{rgb}{0.53,0.81,0.92}
\definecolor{SlateBlue1}{rgb}{0.51,0.44,1.00}
\definecolor{SlateBlue2}{rgb}{0.48,0.40,0.93}
\definecolor{SlateBlue3}{rgb}{0.41,0.35,0.80}
\definecolor{SlateBlue4}{rgb}{0.28,0.24,0.55}
\definecolor{SlateBlue}{rgb}{0.42,0.35,0.80}
\definecolor{SlateGray1}{rgb}{0.78,0.89,1.00}
\definecolor{SlateGray2}{rgb}{0.73,0.83,0.93}
\definecolor{SlateGray3}{rgb}{0.62,0.71,0.80}
\definecolor{SlateGray4}{rgb}{0.42,0.48,0.55}
\definecolor{SlateGray}{rgb}{0.44,0.50,0.56}
\definecolor{SlateGrey}{rgb}{0.44,0.50,0.56}
\definecolor{SpringGreen1}{rgb}{0.00,1.00,0.50}
\definecolor{SpringGreen2}{rgb}{0.00,0.93,0.46}
\definecolor{SpringGreen3}{rgb}{0.00,0.80,0.40}
\definecolor{SpringGreen4}{rgb}{0.00,0.55,0.27}
\definecolor{SpringGreen}{rgb}{0.00,1.00,0.50}
\definecolor{SteelBlue1}{rgb}{0.39,0.72,1.00}
\definecolor{SteelBlue2}{rgb}{0.36,0.67,0.93}
\definecolor{SteelBlue3}{rgb}{0.31,0.58,0.80}
\definecolor{SteelBlue4}{rgb}{0.21,0.39,0.55}
\definecolor{SteelBlue}{rgb}{0.27,0.51,0.71}
\definecolor{VioletRed1}{rgb}{1.00,0.24,0.59}
\definecolor{VioletRed2}{rgb}{0.93,0.23,0.55}
\definecolor{VioletRed3}{rgb}{0.80,0.20,0.47}
\definecolor{VioletRed4}{rgb}{0.55,0.13,0.32}
\definecolor{VioletRed}{rgb}{0.82,0.13,0.56}
\definecolor{WhiteSmoke}{rgb}{0.96,0.96,0.96}
\definecolor{YellowGreen}{rgb}{0.60,0.80,0.20}
\definecolor{aliceblue}{rgb}{0.94,0.97,1.00}
\definecolor{antiquewhite}{rgb}{0.98,0.92,0.84}
\definecolor{aquamarine1}{rgb}{0.50,1.00,0.83}
\definecolor{aquamarine2}{rgb}{0.46,0.93,0.78}
\definecolor{aquamarine3}{rgb}{0.40,0.80,0.67}
\definecolor{aquamarine4}{rgb}{0.27,0.55,0.45}
\definecolor{aquamarine}{rgb}{0.50,1.00,0.83}
\definecolor{azure1}{rgb}{0.94,1.00,1.00}
\definecolor{azure2}{rgb}{0.88,0.93,0.93}
\definecolor{azure3}{rgb}{0.76,0.80,0.80}
\definecolor{azure4}{rgb}{0.51,0.55,0.55}
\definecolor{azure}{rgb}{0.94,1.00,1.00}
\definecolor{beige}{rgb}{0.96,0.96,0.86}
\definecolor{bisque1}{rgb}{1.00,0.89,0.77}
\definecolor{bisque2}{rgb}{0.93,0.84,0.72}
\definecolor{bisque3}{rgb}{0.80,0.72,0.62}
\definecolor{bisque4}{rgb}{0.55,0.49,0.42}
\definecolor{bisque}{rgb}{1.00,0.89,0.77}
\definecolor{black}{rgb}{0.00,0.00,0.00}
\definecolor{blanchedalmond}{rgb}{1.00,0.92,0.80}
\definecolor{blue1}{rgb}{0.00,0.00,1.00}
\definecolor{blue2}{rgb}{0.00,0.00,0.93}
\definecolor{blue3}{rgb}{0.00,0.00,0.80}
\definecolor{blue4}{rgb}{0.00,0.00,0.55}
\definecolor{blueviolet}{rgb}{0.54,0.17,0.89}
\definecolor{blue}{rgb}{0.00,0.00,1.00}
\definecolor{brown1}{rgb}{1.00,0.25,0.25}
\definecolor{brown2}{rgb}{0.93,0.23,0.23}
\definecolor{brown3}{rgb}{0.80,0.20,0.20}
\definecolor{brown4}{rgb}{0.55,0.14,0.14}
\definecolor{brown}{rgb}{0.65,0.16,0.16}
\definecolor{burlywood1}{rgb}{1.00,0.83,0.61}
\definecolor{burlywood2}{rgb}{0.93,0.77,0.57}
\definecolor{burlywood3}{rgb}{0.80,0.67,0.49}
\definecolor{burlywood4}{rgb}{0.55,0.45,0.33}
\definecolor{burlywood}{rgb}{0.87,0.72,0.53}
\definecolor{cadetblue}{rgb}{0.37,0.62,0.63}
\definecolor{chartreuse1}{rgb}{0.50,1.00,0.00}
\definecolor{chartreuse2}{rgb}{0.46,0.93,0.00}
\definecolor{chartreuse3}{rgb}{0.40,0.80,0.00}
\definecolor{chartreuse4}{rgb}{0.27,0.55,0.00}
\definecolor{chartreuse}{rgb}{0.50,1.00,0.00}
\definecolor{chocolate1}{rgb}{1.00,0.50,0.14}
\definecolor{chocolate2}{rgb}{0.93,0.46,0.13}
\definecolor{chocolate3}{rgb}{0.80,0.40,0.11}
\definecolor{chocolate4}{rgb}{0.55,0.27,0.07}
\definecolor{chocolate}{rgb}{0.82,0.41,0.12}
\definecolor{coral1}{rgb}{1.00,0.45,0.34}
\definecolor{coral2}{rgb}{0.93,0.42,0.31}
\definecolor{coral3}{rgb}{0.80,0.36,0.27}
\definecolor{coral4}{rgb}{0.55,0.24,0.18}
\definecolor{coral}{rgb}{1.00,0.50,0.31}
\definecolor{cornflowerblue}{rgb}{0.39,0.58,0.93}
\definecolor{cornsilk1}{rgb}{1.00,0.97,0.86}
\definecolor{cornsilk2}{rgb}{0.93,0.91,0.80}
\definecolor{cornsilk3}{rgb}{0.80,0.78,0.69}
\definecolor{cornsilk4}{rgb}{0.55,0.53,0.47}
\definecolor{cornsilk}{rgb}{1.00,0.97,0.86}
\definecolor{cyan1}{rgb}{0.00,1.00,1.00}
\definecolor{cyan2}{rgb}{0.00,0.93,0.93}
\definecolor{cyan3}{rgb}{0.00,0.80,0.80}
\definecolor{cyan4}{rgb}{0.00,0.55,0.55}
\definecolor{cyan}{rgb}{0.00,1.00,1.00}
\definecolor{darkblue}{rgb}{0.00,0.00,0.55}
\definecolor{darkcyan}{rgb}{0.00,0.55,0.55}
\definecolor{darkgoldenrod}{rgb}{0.72,0.53,0.04}
\definecolor{darkgray}{rgb}{0.66,0.66,0.66}
\definecolor{darkgreen}{rgb}{0.00,0.39,0.00}
\definecolor{darkgrey}{rgb}{0.66,0.66,0.66}
\definecolor{darkkhaki}{rgb}{0.74,0.72,0.42}
\definecolor{darkmagenta}{rgb}{0.55,0.00,0.55}
\definecolor{darkolive}{rgb}{0.33,0.42,0.18}
\definecolor{darkorange}{rgb}{1.00,0.55,0.00}
\definecolor{darkorchid}{rgb}{0.60,0.20,0.80}
\definecolor{darkred}{rgb}{0.55,0.00,0.00}
\definecolor{darksalmon}{rgb}{0.91,0.59,0.48}
\definecolor{darksea}{rgb}{0.56,0.74,0.56}
\definecolor{darkslate}{rgb}{0.18,0.31,0.31}
\definecolor{darkslate}{rgb}{0.18,0.31,0.31}
\definecolor{darkslate}{rgb}{0.28,0.24,0.55}
\definecolor{darkturquoise}{rgb}{0.00,0.81,0.82}
\definecolor{darkviolet}{rgb}{0.58,0.00,0.83}
\definecolor{deeppink}{rgb}{1.00,0.08,0.58}
\definecolor{deepsky}{rgb}{0.00,0.75,1.00}
\definecolor{dimgray}{rgb}{0.41,0.41,0.41}
\definecolor{dimgrey}{rgb}{0.41,0.41,0.41}
\definecolor{dodgerblue}{rgb}{0.12,0.56,1.00}
\definecolor{firebrick1}{rgb}{1.00,0.19,0.19}
\definecolor{firebrick2}{rgb}{0.93,0.17,0.17}
\definecolor{firebrick3}{rgb}{0.80,0.15,0.15}
\definecolor{firebrick4}{rgb}{0.55,0.10,0.10}
\definecolor{firebrick}{rgb}{0.70,0.13,0.13}
\definecolor{floralwhite}{rgb}{1.00,0.98,0.94}
\definecolor{forestgreen}{rgb}{0.13,0.55,0.13}
\definecolor{gainsboro}{rgb}{0.86,0.86,0.86}
\definecolor{ghostwhite}{rgb}{0.97,0.97,1.00}
\definecolor{gold1}{rgb}{1.00,0.84,0.00}
\definecolor{gold2}{rgb}{0.93,0.79,0.00}
\definecolor{gold3}{rgb}{0.80,0.68,0.00}
\definecolor{gold4}{rgb}{0.55,0.46,0.00}
\definecolor{goldenrod1}{rgb}{1.00,0.76,0.15}
\definecolor{goldenrod2}{rgb}{0.93,0.71,0.13}
\definecolor{goldenrod3}{rgb}{0.80,0.61,0.11}
\definecolor{goldenrod4}{rgb}{0.55,0.41,0.08}
\definecolor{goldenrod}{rgb}{0.85,0.65,0.13}
\definecolor{gold}{rgb}{1.00,0.84,0.00}
\definecolor{gray0}{rgb}{0.00,0.00,0.00}
\definecolor{gray100}{rgb}{1.00,1.00,1.00}
\definecolor{gray10}{rgb}{0.10,0.10,0.10}
\definecolor{gray11}{rgb}{0.11,0.11,0.11}
\definecolor{gray12}{rgb}{0.12,0.12,0.12}
\definecolor{gray13}{rgb}{0.13,0.13,0.13}
\definecolor{gray14}{rgb}{0.14,0.14,0.14}
\definecolor{gray15}{rgb}{0.15,0.15,0.15}
\definecolor{gray16}{rgb}{0.16,0.16,0.16}
\definecolor{gray17}{rgb}{0.17,0.17,0.17}
\definecolor{gray18}{rgb}{0.18,0.18,0.18}
\definecolor{gray19}{rgb}{0.19,0.19,0.19}
\definecolor{gray1}{rgb}{0.01,0.01,0.01}
\definecolor{gray20}{rgb}{0.20,0.20,0.20}
\definecolor{gray21}{rgb}{0.21,0.21,0.21}
\definecolor{gray22}{rgb}{0.22,0.22,0.22}
\definecolor{gray23}{rgb}{0.23,0.23,0.23}
\definecolor{gray24}{rgb}{0.24,0.24,0.24}
\definecolor{gray25}{rgb}{0.25,0.25,0.25}
\definecolor{gray26}{rgb}{0.26,0.26,0.26}
\definecolor{gray27}{rgb}{0.27,0.27,0.27}
\definecolor{gray28}{rgb}{0.28,0.28,0.28}
\definecolor{gray29}{rgb}{0.29,0.29,0.29}
\definecolor{gray2}{rgb}{0.02,0.02,0.02}
\definecolor{gray30}{rgb}{0.30,0.30,0.30}
\definecolor{gray31}{rgb}{0.31,0.31,0.31}
\definecolor{gray32}{rgb}{0.32,0.32,0.32}
\definecolor{gray33}{rgb}{0.33,0.33,0.33}
\definecolor{gray34}{rgb}{0.34,0.34,0.34}
\definecolor{gray35}{rgb}{0.35,0.35,0.35}
\definecolor{gray36}{rgb}{0.36,0.36,0.36}
\definecolor{gray37}{rgb}{0.37,0.37,0.37}
\definecolor{gray38}{rgb}{0.38,0.38,0.38}
\definecolor{gray39}{rgb}{0.39,0.39,0.39}
\definecolor{gray3}{rgb}{0.03,0.03,0.03}
\definecolor{gray40}{rgb}{0.40,0.40,0.40}
\definecolor{gray41}{rgb}{0.41,0.41,0.41}
\definecolor{gray42}{rgb}{0.42,0.42,0.42}
\definecolor{gray43}{rgb}{0.43,0.43,0.43}
\definecolor{gray44}{rgb}{0.44,0.44,0.44}
\definecolor{gray45}{rgb}{0.45,0.45,0.45}
\definecolor{gray46}{rgb}{0.46,0.46,0.46}
\definecolor{gray47}{rgb}{0.47,0.47,0.47}
\definecolor{gray48}{rgb}{0.48,0.48,0.48}
\definecolor{gray49}{rgb}{0.49,0.49,0.49}
\definecolor{gray4}{rgb}{0.04,0.04,0.04}
\definecolor{gray50}{rgb}{0.50,0.50,0.50}
\definecolor{gray51}{rgb}{0.51,0.51,0.51}
\definecolor{gray52}{rgb}{0.52,0.52,0.52}
\definecolor{gray53}{rgb}{0.53,0.53,0.53}
\definecolor{gray54}{rgb}{0.54,0.54,0.54}
\definecolor{gray55}{rgb}{0.55,0.55,0.55}
\definecolor{gray56}{rgb}{0.56,0.56,0.56}
\definecolor{gray57}{rgb}{0.57,0.57,0.57}
\definecolor{gray58}{rgb}{0.58,0.58,0.58}
\definecolor{gray59}{rgb}{0.59,0.59,0.59}
\definecolor{gray5}{rgb}{0.05,0.05,0.05}
\definecolor{gray60}{rgb}{0.60,0.60,0.60}
\definecolor{gray61}{rgb}{0.61,0.61,0.61}
\definecolor{gray62}{rgb}{0.62,0.62,0.62}
\definecolor{gray63}{rgb}{0.63,0.63,0.63}
\definecolor{gray64}{rgb}{0.64,0.64,0.64}
\definecolor{gray65}{rgb}{0.65,0.65,0.65}
\definecolor{gray66}{rgb}{0.66,0.66,0.66}
\definecolor{gray67}{rgb}{0.67,0.67,0.67}
\definecolor{gray68}{rgb}{0.68,0.68,0.68}
\definecolor{gray69}{rgb}{0.69,0.69,0.69}
\definecolor{gray6}{rgb}{0.06,0.06,0.06}
\definecolor{gray70}{rgb}{0.70,0.70,0.70}
\definecolor{gray71}{rgb}{0.71,0.71,0.71}
\definecolor{gray72}{rgb}{0.72,0.72,0.72}
\definecolor{gray73}{rgb}{0.73,0.73,0.73}
\definecolor{gray74}{rgb}{0.74,0.74,0.74}
\definecolor{gray75}{rgb}{0.75,0.75,0.75}
\definecolor{gray76}{rgb}{0.76,0.76,0.76}
\definecolor{gray77}{rgb}{0.77,0.77,0.77}
\definecolor{gray78}{rgb}{0.78,0.78,0.78}
\definecolor{gray79}{rgb}{0.79,0.79,0.79}
\definecolor{gray7}{rgb}{0.07,0.07,0.07}
\definecolor{gray80}{rgb}{0.80,0.80,0.80}
\definecolor{gray81}{rgb}{0.81,0.81,0.81}
\definecolor{gray82}{rgb}{0.82,0.82,0.82}
\definecolor{gray83}{rgb}{0.83,0.83,0.83}
\definecolor{gray84}{rgb}{0.84,0.84,0.84}
\definecolor{gray85}{rgb}{0.85,0.85,0.85}
\definecolor{gray86}{rgb}{0.86,0.86,0.86}
\definecolor{gray87}{rgb}{0.87,0.87,0.87}
\definecolor{gray88}{rgb}{0.88,0.88,0.88}
\definecolor{gray89}{rgb}{0.89,0.89,0.89}
\definecolor{gray8}{rgb}{0.08,0.08,0.08}
\definecolor{gray90}{rgb}{0.90,0.90,0.90}
\definecolor{gray91}{rgb}{0.91,0.91,0.91}
\definecolor{gray92}{rgb}{0.92,0.92,0.92}
\definecolor{gray93}{rgb}{0.93,0.93,0.93}
\definecolor{gray94}{rgb}{0.94,0.94,0.94}
\definecolor{gray95}{rgb}{0.95,0.95,0.95}
\definecolor{gray96}{rgb}{0.96,0.96,0.96}
\definecolor{gray97}{rgb}{0.97,0.97,0.97}
\definecolor{gray98}{rgb}{0.98,0.98,0.98}
\definecolor{gray99}{rgb}{0.99,0.99,0.99}
\definecolor{gray9}{rgb}{0.09,0.09,0.09}
\definecolor{gray}{rgb}{0.75,0.75,0.75}
\definecolor{green1}{rgb}{0.00,1.00,0.00}
\definecolor{green2}{rgb}{0.00,0.93,0.00}
\definecolor{green3}{rgb}{0.00,0.80,0.00}
\definecolor{green4}{rgb}{0.00,0.55,0.00}
\definecolor{greenyellow}{rgb}{0.68,1.00,0.18}
\definecolor{green}{rgb}{0.00,1.00,0.00}
\definecolor{grey0}{rgb}{0.00,0.00,0.00}
\definecolor{grey100}{rgb}{1.00,1.00,1.00}
\definecolor{grey10}{rgb}{0.10,0.10,0.10}
\definecolor{grey11}{rgb}{0.11,0.11,0.11}
\definecolor{grey12}{rgb}{0.12,0.12,0.12}
\definecolor{grey13}{rgb}{0.13,0.13,0.13}
\definecolor{grey14}{rgb}{0.14,0.14,0.14}
\definecolor{grey15}{rgb}{0.15,0.15,0.15}
\definecolor{grey16}{rgb}{0.16,0.16,0.16}
\definecolor{grey17}{rgb}{0.17,0.17,0.17}
\definecolor{grey18}{rgb}{0.18,0.18,0.18}
\definecolor{grey19}{rgb}{0.19,0.19,0.19}
\definecolor{grey1}{rgb}{0.01,0.01,0.01}
\definecolor{grey20}{rgb}{0.20,0.20,0.20}
\definecolor{grey21}{rgb}{0.21,0.21,0.21}
\definecolor{grey22}{rgb}{0.22,0.22,0.22}
\definecolor{grey23}{rgb}{0.23,0.23,0.23}
\definecolor{grey24}{rgb}{0.24,0.24,0.24}
\definecolor{grey25}{rgb}{0.25,0.25,0.25}
\definecolor{grey26}{rgb}{0.26,0.26,0.26}
\definecolor{grey27}{rgb}{0.27,0.27,0.27}
\definecolor{grey28}{rgb}{0.28,0.28,0.28}
\definecolor{grey29}{rgb}{0.29,0.29,0.29}
\definecolor{grey2}{rgb}{0.02,0.02,0.02}
\definecolor{grey30}{rgb}{0.30,0.30,0.30}
\definecolor{grey31}{rgb}{0.31,0.31,0.31}
\definecolor{grey32}{rgb}{0.32,0.32,0.32}
\definecolor{grey33}{rgb}{0.33,0.33,0.33}
\definecolor{grey34}{rgb}{0.34,0.34,0.34}
\definecolor{grey35}{rgb}{0.35,0.35,0.35}
\definecolor{grey36}{rgb}{0.36,0.36,0.36}
\definecolor{grey37}{rgb}{0.37,0.37,0.37}
\definecolor{grey38}{rgb}{0.38,0.38,0.38}
\definecolor{grey39}{rgb}{0.39,0.39,0.39}
\definecolor{grey3}{rgb}{0.03,0.03,0.03}
\definecolor{grey40}{rgb}{0.40,0.40,0.40}
\definecolor{grey41}{rgb}{0.41,0.41,0.41}
\definecolor{grey42}{rgb}{0.42,0.42,0.42}
\definecolor{grey43}{rgb}{0.43,0.43,0.43}
\definecolor{grey44}{rgb}{0.44,0.44,0.44}
\definecolor{grey45}{rgb}{0.45,0.45,0.45}
\definecolor{grey46}{rgb}{0.46,0.46,0.46}
\definecolor{grey47}{rgb}{0.47,0.47,0.47}
\definecolor{grey48}{rgb}{0.48,0.48,0.48}
\definecolor{grey49}{rgb}{0.49,0.49,0.49}
\definecolor{grey4}{rgb}{0.04,0.04,0.04}
\definecolor{grey50}{rgb}{0.50,0.50,0.50}
\definecolor{grey51}{rgb}{0.51,0.51,0.51}
\definecolor{grey52}{rgb}{0.52,0.52,0.52}
\definecolor{grey53}{rgb}{0.53,0.53,0.53}
\definecolor{grey54}{rgb}{0.54,0.54,0.54}
\definecolor{grey55}{rgb}{0.55,0.55,0.55}
\definecolor{grey56}{rgb}{0.56,0.56,0.56}
\definecolor{grey57}{rgb}{0.57,0.57,0.57}
\definecolor{grey58}{rgb}{0.58,0.58,0.58}
\definecolor{grey59}{rgb}{0.59,0.59,0.59}
\definecolor{grey5}{rgb}{0.05,0.05,0.05}
\definecolor{grey60}{rgb}{0.60,0.60,0.60}
\definecolor{grey61}{rgb}{0.61,0.61,0.61}
\definecolor{grey62}{rgb}{0.62,0.62,0.62}
\definecolor{grey63}{rgb}{0.63,0.63,0.63}
\definecolor{grey64}{rgb}{0.64,0.64,0.64}
\definecolor{grey65}{rgb}{0.65,0.65,0.65}
\definecolor{grey66}{rgb}{0.66,0.66,0.66}
\definecolor{grey67}{rgb}{0.67,0.67,0.67}
\definecolor{grey68}{rgb}{0.68,0.68,0.68}
\definecolor{grey69}{rgb}{0.69,0.69,0.69}
\definecolor{grey6}{rgb}{0.06,0.06,0.06}
\definecolor{grey70}{rgb}{0.70,0.70,0.70}
\definecolor{grey71}{rgb}{0.71,0.71,0.71}
\definecolor{grey72}{rgb}{0.72,0.72,0.72}
\definecolor{grey73}{rgb}{0.73,0.73,0.73}
\definecolor{grey74}{rgb}{0.74,0.74,0.74}
\definecolor{grey75}{rgb}{0.75,0.75,0.75}
\definecolor{grey76}{rgb}{0.76,0.76,0.76}
\definecolor{grey77}{rgb}{0.77,0.77,0.77}
\definecolor{grey78}{rgb}{0.78,0.78,0.78}
\definecolor{grey79}{rgb}{0.79,0.79,0.79}
\definecolor{grey7}{rgb}{0.07,0.07,0.07}
\definecolor{grey80}{rgb}{0.80,0.80,0.80}
\definecolor{grey81}{rgb}{0.81,0.81,0.81}
\definecolor{grey82}{rgb}{0.82,0.82,0.82}
\definecolor{grey83}{rgb}{0.83,0.83,0.83}
\definecolor{grey84}{rgb}{0.84,0.84,0.84}
\definecolor{grey85}{rgb}{0.85,0.85,0.85}
\definecolor{grey86}{rgb}{0.86,0.86,0.86}
\definecolor{grey87}{rgb}{0.87,0.87,0.87}
\definecolor{grey88}{rgb}{0.88,0.88,0.88}
\definecolor{grey89}{rgb}{0.89,0.89,0.89}
\definecolor{grey8}{rgb}{0.08,0.08,0.08}
\definecolor{grey90}{rgb}{0.90,0.90,0.90}
\definecolor{grey91}{rgb}{0.91,0.91,0.91}
\definecolor{grey92}{rgb}{0.92,0.92,0.92}
\definecolor{grey93}{rgb}{0.93,0.93,0.93}
\definecolor{grey94}{rgb}{0.94,0.94,0.94}
\definecolor{grey95}{rgb}{0.95,0.95,0.95}
\definecolor{grey96}{rgb}{0.96,0.96,0.96}
\definecolor{grey97}{rgb}{0.97,0.97,0.97}
\definecolor{grey98}{rgb}{0.98,0.98,0.98}
\definecolor{grey99}{rgb}{0.99,0.99,0.99}
\definecolor{grey9}{rgb}{0.09,0.09,0.09}
\definecolor{grey}{rgb}{0.75,0.75,0.75}
\definecolor{honeydew1}{rgb}{0.94,1.00,0.94}
\definecolor{honeydew2}{rgb}{0.88,0.93,0.88}
\definecolor{honeydew3}{rgb}{0.76,0.80,0.76}
\definecolor{honeydew4}{rgb}{0.51,0.55,0.51}
\definecolor{honeydew}{rgb}{0.94,1.00,0.94}
\definecolor{hotpink}{rgb}{1.00,0.41,0.71}
\definecolor{indianred}{rgb}{0.80,0.36,0.36}
\definecolor{ivory1}{rgb}{1.00,1.00,0.94}
\definecolor{ivory2}{rgb}{0.93,0.93,0.88}
\definecolor{ivory3}{rgb}{0.80,0.80,0.76}
\definecolor{ivory4}{rgb}{0.55,0.55,0.51}
\definecolor{ivory}{rgb}{1.00,1.00,0.94}
\definecolor{khaki1}{rgb}{1.00,0.96,0.56}
\definecolor{khaki2}{rgb}{0.93,0.90,0.52}
\definecolor{khaki3}{rgb}{0.80,0.78,0.45}
\definecolor{khaki4}{rgb}{0.55,0.53,0.31}
\definecolor{khaki}{rgb}{0.94,0.90,0.55}
\definecolor{lavenderblush}{rgb}{1.00,0.94,0.96}
\definecolor{lavender}{rgb}{0.90,0.90,0.98}
\definecolor{lawngreen}{rgb}{0.49,0.99,0.00}
\definecolor{lemonchiffon}{rgb}{1.00,0.98,0.80}
\definecolor{lightblue}{rgb}{0.68,0.85,0.90}
\definecolor{lightcoral}{rgb}{0.94,0.50,0.50}
\definecolor{lightcyan}{rgb}{0.88,1.00,1.00}
\definecolor{lightgoldenrod}{rgb}{0.93,0.87,0.51}
\definecolor{lightgoldenrod}{rgb}{0.98,0.98,0.82}
\definecolor{lightgray}{rgb}{0.83,0.83,0.83}
\definecolor{lightgreen}{rgb}{0.56,0.93,0.56}
\definecolor{lightgrey}{rgb}{0.83,0.83,0.83}
\definecolor{lightpink}{rgb}{1.00,0.71,0.76}
\definecolor{lightsalmon}{rgb}{1.00,0.63,0.48}
\definecolor{lightsea}{rgb}{0.13,0.70,0.67}
\definecolor{lightsky}{rgb}{0.53,0.81,0.98}
\definecolor{lightslate}{rgb}{0.47,0.53,0.60}
\definecolor{lightslate}{rgb}{0.47,0.53,0.60}
\definecolor{lightslate}{rgb}{0.52,0.44,1.00}
\definecolor{lightsteel}{rgb}{0.69,0.77,0.87}
\definecolor{lightyellow}{rgb}{1.00,1.00,0.88}
\definecolor{limegreen}{rgb}{0.20,0.80,0.20}
\definecolor{linen}{rgb}{0.98,0.94,0.90}
\definecolor{magenta1}{rgb}{1.00,0.00,1.00}
\definecolor{magenta2}{rgb}{0.93,0.00,0.93}
\definecolor{magenta3}{rgb}{0.80,0.00,0.80}
\definecolor{magenta4}{rgb}{0.55,0.00,0.55}
\definecolor{magenta}{rgb}{1.00,0.00,1.00}
\definecolor{maroon1}{rgb}{1.00,0.20,0.70}
\definecolor{maroon2}{rgb}{0.93,0.19,0.65}
\definecolor{maroon3}{rgb}{0.80,0.16,0.56}
\definecolor{maroon4}{rgb}{0.55,0.11,0.38}
\definecolor{maroon}{rgb}{0.69,0.19,0.38}
\definecolor{mediumaquamarine}{rgb}{0.40,0.80,0.67}
\definecolor{mediumblue}{rgb}{0.00,0.00,0.80}
\definecolor{mediumorchid}{rgb}{0.73,0.33,0.83}
\definecolor{mediumpurple}{rgb}{0.58,0.44,0.86}
\definecolor{mediumsea}{rgb}{0.24,0.70,0.44}
\definecolor{mediumslate}{rgb}{0.48,0.41,0.93}
\definecolor{mediumspring}{rgb}{0.00,0.98,0.60}
\definecolor{mediumturquoise}{rgb}{0.28,0.82,0.80}
\definecolor{mediumviolet}{rgb}{0.78,0.08,0.52}
\definecolor{midnightblue}{rgb}{0.10,0.10,0.44}
\definecolor{mintcream}{rgb}{0.96,1.00,0.98}
\definecolor{mistyrose}{rgb}{1.00,0.89,0.88}
\definecolor{moccasin}{rgb}{1.00,0.89,0.71}
\definecolor{navajowhite}{rgb}{1.00,0.87,0.68}
\definecolor{navyblue}{rgb}{0.00,0.00,0.50}
\definecolor{navy}{rgb}{0.00,0.00,0.50}
\definecolor{oldlace}{rgb}{0.99,0.96,0.90}
\definecolor{olivedrab}{rgb}{0.42,0.56,0.14}
\definecolor{orange1}{rgb}{1.00,0.65,0.00}
\definecolor{orange2}{rgb}{0.93,0.60,0.00}
\definecolor{orange3}{rgb}{0.80,0.52,0.00}
\definecolor{orange4}{rgb}{0.55,0.35,0.00}
\definecolor{orangered}{rgb}{1.00,0.27,0.00}
\definecolor{orange}{rgb}{1.00,0.65,0.00}
\definecolor{orchid1}{rgb}{1.00,0.51,0.98}
\definecolor{orchid2}{rgb}{0.93,0.48,0.91}
\definecolor{orchid3}{rgb}{0.80,0.41,0.79}
\definecolor{orchid4}{rgb}{0.55,0.28,0.54}
\definecolor{orchid}{rgb}{0.85,0.44,0.84}
\definecolor{palegoldenrod}{rgb}{0.93,0.91,0.67}
\definecolor{palegreen}{rgb}{0.60,0.98,0.60}
\definecolor{paleturquoise}{rgb}{0.69,0.93,0.93}
\definecolor{paleviolet}{rgb}{0.86,0.44,0.58}
\definecolor{papayawhip}{rgb}{1.00,0.94,0.84}
\definecolor{peachpuff}{rgb}{1.00,0.85,0.73}
\definecolor{peru}{rgb}{0.80,0.52,0.25}
\definecolor{pink1}{rgb}{1.00,0.71,0.77}
\definecolor{pink2}{rgb}{0.93,0.66,0.72}
\definecolor{pink3}{rgb}{0.80,0.57,0.62}
\definecolor{pink4}{rgb}{0.55,0.39,0.42}
\definecolor{pink}{rgb}{1.00,0.75,0.80}
\definecolor{plum1}{rgb}{1.00,0.73,1.00}
\definecolor{plum2}{rgb}{0.93,0.68,0.93}
\definecolor{plum3}{rgb}{0.80,0.59,0.80}
\definecolor{plum4}{rgb}{0.55,0.40,0.55}
\definecolor{plum}{rgb}{0.87,0.63,0.87}
\definecolor{powderblue}{rgb}{0.69,0.88,0.90}
\definecolor{purple1}{rgb}{0.61,0.19,1.00}
\definecolor{purple2}{rgb}{0.57,0.17,0.93}
\definecolor{purple3}{rgb}{0.49,0.15,0.80}
\definecolor{purple4}{rgb}{0.33,0.10,0.55}
\definecolor{purple}{rgb}{0.63,0.13,0.94}
\definecolor{red1}{rgb}{1.00,0.00,0.00}
\definecolor{red2}{rgb}{0.93,0.00,0.00}
\definecolor{red3}{rgb}{0.80,0.00,0.00}
\definecolor{red4}{rgb}{0.55,0.00,0.00}
\definecolor{red}{rgb}{1.00,0.00,0.00}
\definecolor{rosybrown}{rgb}{0.74,0.56,0.56}
\definecolor{royalblue}{rgb}{0.25,0.41,0.88}
\definecolor{saddlebrown}{rgb}{0.55,0.27,0.07}
\definecolor{salmon1}{rgb}{1.00,0.55,0.41}
\definecolor{salmon2}{rgb}{0.93,0.51,0.38}
\definecolor{salmon3}{rgb}{0.80,0.44,0.33}
\definecolor{salmon4}{rgb}{0.55,0.30,0.22}
\definecolor{salmon}{rgb}{0.98,0.50,0.45}
\definecolor{sandybrown}{rgb}{0.96,0.64,0.38}
\definecolor{seagreen}{rgb}{0.18,0.55,0.34}
\definecolor{seashell1}{rgb}{1.00,0.96,0.93}
\definecolor{seashell2}{rgb}{0.93,0.90,0.87}
\definecolor{seashell3}{rgb}{0.80,0.77,0.75}
\definecolor{seashell4}{rgb}{0.55,0.53,0.51}
\definecolor{seashell}{rgb}{1.00,0.96,0.93}
\definecolor{sienna1}{rgb}{1.00,0.51,0.28}
\definecolor{sienna2}{rgb}{0.93,0.47,0.26}
\definecolor{sienna3}{rgb}{0.80,0.41,0.22}
\definecolor{sienna4}{rgb}{0.55,0.28,0.15}
\definecolor{sienna}{rgb}{0.63,0.32,0.18}
\definecolor{skyblue}{rgb}{0.53,0.81,0.92}
\definecolor{slateblue}{rgb}{0.42,0.35,0.80}
\definecolor{slategray}{rgb}{0.44,0.50,0.56}
\definecolor{slategrey}{rgb}{0.44,0.50,0.56}
\definecolor{snow1}{rgb}{1.00,0.98,0.98}
\definecolor{snow2}{rgb}{0.93,0.91,0.91}
\definecolor{snow3}{rgb}{0.80,0.79,0.79}
\definecolor{snow4}{rgb}{0.55,0.54,0.54}
\definecolor{snow}{rgb}{1.00,0.98,0.98}
\definecolor{springgreen}{rgb}{0.00,1.00,0.50}
\definecolor{steelblue}{rgb}{0.27,0.51,0.71}
\definecolor{tan1}{rgb}{1.00,0.65,0.31}
\definecolor{tan2}{rgb}{0.93,0.60,0.29}
\definecolor{tan3}{rgb}{0.80,0.52,0.25}
\definecolor{tan4}{rgb}{0.55,0.35,0.17}
\definecolor{tan}{rgb}{0.82,0.71,0.55}
\definecolor{thistle1}{rgb}{1.00,0.88,1.00}
\definecolor{thistle2}{rgb}{0.93,0.82,0.93}
\definecolor{thistle3}{rgb}{0.80,0.71,0.80}
\definecolor{thistle4}{rgb}{0.55,0.48,0.55}
\definecolor{thistle}{rgb}{0.85,0.75,0.85}
\definecolor{tomato1}{rgb}{1.00,0.39,0.28}
\definecolor{tomato2}{rgb}{0.93,0.36,0.26}
\definecolor{tomato3}{rgb}{0.80,0.31,0.22}
\definecolor{tomato4}{rgb}{0.55,0.21,0.15}
\definecolor{tomato}{rgb}{1.00,0.39,0.28}
\definecolor{turquoise1}{rgb}{0.00,0.96,1.00}
\definecolor{turquoise2}{rgb}{0.00,0.90,0.93}
\definecolor{turquoise3}{rgb}{0.00,0.77,0.80}
\definecolor{turquoise4}{rgb}{0.00,0.53,0.55}
\definecolor{turquoise}{rgb}{0.25,0.88,0.82}
\definecolor{violetred}{rgb}{0.82,0.13,0.56}
\definecolor{violet}{rgb}{0.93,0.51,0.93}
\definecolor{wheat1}{rgb}{1.00,0.91,0.73}
\definecolor{wheat2}{rgb}{0.93,0.85,0.68}
\definecolor{wheat3}{rgb}{0.80,0.73,0.59}
\definecolor{wheat4}{rgb}{0.55,0.49,0.40}
\definecolor{wheat}{rgb}{0.96,0.87,0.70}
\definecolor{whitesmoke}{rgb}{0.96,0.96,0.96}
\definecolor{white}{rgb}{1.00,1.00,1.00}
\definecolor{yellow1}{rgb}{1.00,1.00,0.00}
\definecolor{yellow2}{rgb}{0.93,0.93,0.00}
\definecolor{yellow3}{rgb}{0.80,0.80,0.00}
\definecolor{yellow4}{rgb}{0.55,0.55,0.00}
\definecolor{yellowgreen}{rgb}{0.60,0.80,0.20}
\definecolor{yellow}{rgb}{1.00,1.00,0.00}

\usepackage{etoolbox}
\usepackage{lscape}
\usepackage{longtable}

\newcommand\suzaku{{\it Suzaku}}
\newcommand\nustar{{\it NuSTAR}}
\newcommand\uhuru{{\it UHURU}}
\newcommand\arielv{{\it ARIEL~V}}
\newcommand\heada{{\it HEAD 1~A1}}
\newcommand\headb{{\it HEAD 1~A2}}
\newcommand\einstein{{\it Einstein}}
\newcommand\exosat{{\it EXOSAT}}
\newcommand\swift{{\it Swift}}
\newcommand\asca{{\it ASCA}}

\newcommand\rxte{{\it RXTE}}

\newcommand\astrosat{{\it ASTROSAT}}
\newcommand\xmm{{\it XMM-Newton}}

\newcommand\s{{\rm~s}}

\newcommand\kev{{\rm~keV}}

\newcommand\xiunit{\ifmmode {\rm~erg\s}$^{-1}$ \else ~erg~cm~s$^{-1}$\fi}
\newcommand\kms{\ifmmode {\rm~km\ s}$^{-1}$ \else ~km s$^{-1}$\fi}
\newcommand\Hunit{\ifmmode {\rm~km\ s}$^{-1}$\ {\rm Mpc}$^{-1}$
        \else ~km s$^{-1}$ Mpc$^{-1}$\fi}
\newcommand\ctssec{\ifmmode {\rm~count\ s}$^{-1}$ \else ~count s$^{-1}$\fi}
\newcommand\ergsec{\ifmmode {\rm~erg\ s}$^{-1}$ \else
        ~erg s$^{-1}$\fi}
\newcommand\funit{\ifmmode {\rm~erg\ s}$^{-1}$\;{\rm cm}$^{-2}$ \else
        ~ergs s$^{-1}$ cm$^{-2}$\fi}
\newcommand\phflux{\ifmmode {\rm~photon\ s}$^{-1}$\;{\rm cm}$^{-i2}$
        \else   ~photon s$^{-1}$ cm$^{-2}$\fi}
\newcommand\efluxA{\ifmmode {\rm~erg\ s}$^{-1}$\;{\rm cm}$^{-2}$\;{\rm
        \AA}$^{-1}$ \else ~erg s$^{-1}$ cm$^{-2}$ \AA$^{-1}$\fi}
\newcommand\efluxHz{\ifmmode {\rm~erg\ s}$^{-1}$\;{\rm cm}$^{-2}$\;{\rm
        Hz}$^{-1}$ \else ~erg s$^{-1}$ cm$^{-2}$ Hz$^{-1}$\fi}
\newcommand\cc{\ifmmode {\rm~cm}$^{-3}$ \else cm$^{-3}$\fi}
\newcommand\FWHM{\ifmmode {\rm~FWHM} \else ${\rm~FWHM}$\fi}
\newcommand\Msun{\ifmmode M_{\odot} \else $M_{\odot}$\fi}
\newcommand\Lsun{\ifmmode L_{\odot} \else $L_{\odot}$\fi}

\newcommand\hbeta{\ifmmode {\rm H}\beta \else H$\beta$\fi}
\newcommand\Kalpha{\ifmmode {\rm K}\alpha \else K$\alpha$\fi}
\newcommand\nh{\ifmmode N_{\rm H} \else N$_{\rm H}$\fi}

\makeatletter

\newcommand{\Rmnum}[1]{\expandafter\@slowromancap\romannumeral #1@}
\makeatother

\title [Correlated variability of Fairall~9] {Correlated X-ray/UV/optical variability and the nature of accretion disc in  the bare Seyfert 1 galaxy Fairall~9}

\author [Pal et al.] { Main Pal$^{1}$\thanks { Email: mainpal@iucaa.in},
Gulab \ C.\ Dewangan $^{1}$\footnotemark[1], S.~D.~Connolly $^{2}$\footnotemark[2] \& Ranjeev Misra $^{1}$ \footnotemark[1]\\
$^{1}$Inter University Centre for Astronomy and Astrophysics (IUCAA), Pune 411007, India.\\
$^{2}$Physics \& Astronomy, University of Southampton, Highfield, Southampton, SO17~1BJ, UK.}

\begin{document}

\pagerange{\pageref{firstpage}--\pageref{lastpage}} \pubyear{}
\date{\today}
\maketitle
\begin{abstract}

 We study multi-wavelength variability of a bare Seyfert 1 galaxy Fairall~9 using  \swift{} monitoring observations consisting of $165$ usable pointings spanning nearly two years and covering six UV/optical bands and X-rays.  Fairall~9 is highly variable in all bands though the variability amplitude decreases from X-ray to optical bands. The variations in the X-ray and UV/optical bands are strongly correlated. Our reverberation mapping analysis using the {\tt JAVALIN} tool shows that the variation in the UV/optical bands lag behind the variations in the X-ray band by $\sim 2-10{\rm~days}$. These lag measurements strongly suggest that the optical/UV variations are mainly caused by variations in the X-rays, and the origin of most of the optical/UV emission is  X-ray reprocessing. The observed lags are found to vary as $\tau\propto\lambda^{1.36\pm0.13}$, consistent with the prediction, $\tau\propto\lambda^{4/3}$, for X-ray reprocessing in a standard accretion disc. However, the predicted lags for an standard accretion disc with X-ray reprocessing using black hole mass ($M_{BH} \sim 2.6\times10^{8}~M_{\odot}$) estimated from the reverberation mapping of broad emission lines and accretion rate relative to the Eddington rate ($\dot{m_E} =0.02$) are shorter than the observed lags. These observations suggest that accretion disc in Fairall~9 is larger than that predicted by the standard disc model, and confirm similar findings in a few other Seyfert 1 galaxies such as NGC~5548.

\end{abstract}


\section{Introduction}
Active galactic nuclei (AGN) are thought to contain an accreting supermassive black hole (SMBH) at the centre of the parent galaxy. Such sources emit over the entire electromagnetic spectrum from radio to X-rays. The main radiative component in the X-ray band is the power-law emission which is thought to arise due to the Compton upscattering of seed photons in an optically thin and hot corona \citep{1980A&A....86..121S,1991ApJ...380L..51H}. The seed photons are thought to be associated with the material accreting towards the SMBH forming an accretion disc. The popular standard accretion disc model, which describes an optically thick and geometrically thin disc, was proposed by \citet{1973A&A....24..337S}. %Half of the gravitational potential energy released by the accreted material is dissipated through viscous heating in the accretion disc. 
The dissipated energy in the accretion disc is considered to be radiated away locally as blackbody emission at each radius of the accretion disc. The temperature at each radius varies inversely as a three--fourths power of the radial distance from the centre. The emission from outer to inner regions of the accretion disc is observed in the optical to UV bands (e. g., \citealt{1999PASP..111....1K}) and sometimes partly at the softest X-rays (e.g., \citealt{1996A&A...305...53B}).

 The emission from the accretion disc in Seyfert galaxies peaks in the UV regime where the peak of the UV emission is hard to detect due to Galactic absorption along the line of sight. The observed UV/optical emission show a variety of variability on various timescales from hours to years for a range of black hole masses (e.g., MR~2251-178: \citealt{2008MNRAS.389.1479A}, Mrk~509: \citealt{2011A&A...534A..39M}, NGC~5548: \citealt{2014MNRAS.444.1469M}, NGC~2617: \citealt{2014ApJ...788...48S}, NGC~4395: \citealt{2016arXiv160100215M}). However, the origin of the disc emission variability is less well understood. One can assume that the natural origin of the UV/optical emission variability is associated with the variations in the accretion flow (e.g., \citealt{2008ApJ...677..880M,2008MNRAS.389.1479A}). Such models pose questions on the explanation of the observed disc variability. Sometimes the timescale of observed fluctuations is of the order of hours to days which is much faster than the expected timescale of the accretion flow. Another problem is that the changes in the UV/optical emission lag the X-ray emission, which is unexplainable by the above assumption (e.g., \citealt{2014ApJ...788...48S,2014MNRAS.444.1469M}).

 As suggested by \citet{1991ApJ...371..541K}, the variations in UV/optical emission can be delayed with respect to X-rays if the variations in the UV/optical emission follow the changes occurring in the X-ray emission. This can happen if X-rays get absorbed in an optically thick medium such as the accretion disc and reprocess into optical/UV emission. For X-ray reprocessing in a standard optically thick and geometrically thin accretion disc, the predicted delay in the UV/optical bands scales as $\tau\propto\lambda^{4/3}$ \citep{1999MNRAS.302L..24C}. The observed lags in the multi-wavelength study of several AGN have been found consistent with this prediction of X-ray reprocessing \citep{2007MNRAS.380..669C,2014MNRAS.444.1469M,2016arXiv160100215M,2016MNRAS.456.4040T,2015ApJ...806..129E,2015arXiv151005648F}. However, \citet{2008RMxAC..32....1G} has argued that the variations in the UV/optical emission are independent from the variations of X-rays and suggested that the fluctuations are produced locally. Furthermore, in a Seyfert galaxy NGC~3783, \citet{2009MNRAS.397.2004A} found more rapid variations in optical bands than those of the X-rays indicating distinct regions for their origin.
 
 Multi-wavelength monitoring of AGN provides a window to study the various regions of the central regions e.g., accretion discs. Before the era of \xmm{}~and \swift{}, only a few simultaneous long and intensive multi-wavelength campaigns were undertaken. Monitoring campaigns with \rxte{} and ground-based optical telescopes were performed on a small number of AGN (e.g., \citealt{2000ApJ...534..180E,2009MNRAS.397.2004A,2010MNRAS.403..605B}). The space-based \xmm{}~and \swift{}~satellites have the capability to make multi-wavelength observations in the optical/UV and X-ray bands. Multi-wavelength observations of Seyfert 1 galaxies have revealed  correlated variability in the optical/UV and X-ray bands \citep{2000ApJ...542..161P,2003ApJ...584L..53U,2011A&A...534A..39M,2014MNRAS.444.1469M,2015ebha.confE..49M,2016MNRAS.456.4040T,2015ApJ...806..129E,
2015arXiv151005648F,2016arXiv160508050N}. However, some studies on Seyfert galaxies have shown a relatively moderate correlation between the UV/optical and X-rays \citep[e.g., NGC~7469:][]{1998ApJ...505..594N} and one of the Seyfert NGC~3516 showed no clear relationship between optical and X-rays \citep{2002AJ....124.1988M}. Thus, the variations and correlation between the UV/optical and X-rays are complex and may depend on a number of factors. To study the multi-band variability,  the Seyfert 1 galaxy Fairall~9 is an ideal source which is well known to possess a bare nucleus without any significant intrinsic neutral or warm absorption \citep{2001A&A...373..805G, 2011MNRAS.415.1895E,2016arXiv160205589L}.

Fairall 9 \citep{1977MNRAS.180..391F} is a radio--quiet Seyfert type 1 AGN located at $z=0.0461$ \citep{1989spce.book.....L}. The estimated mass of the black hole is $2.55\pm0.56~\times10^{8}~M_{\odot}$ based on reverberation mapping \citep{2004ApJ...613..682P}.
Since its discovery, it has been observed extensively in the X-ray bands (\uhuru{} : \citealt{1978ApJS...38..357F}, \arielv{} : \citealt{1981MNRAS.197..893M}, \heada{} : \citealt{1984ApJS...56..507W}, \headb{}: \citealt{1982ApJ...253..485P}, \einstein{} : \citealt{1984ApJ...280..499P}, \exosat{} : \citealt{1986ApJ...307..486M}, \asca{} : \citealt{1997MNRAS.286..513R}, \xmm{} : \citealt{2001A&A...373..805G,2011MNRAS.415.1895E}, \suzaku{} : \citealt{2009ApJ...703.2171S}). Detailed studies of the broadband X-ray spectrum have revealed no signatures of warm absorbers at low energies making this AGN to possess a bare nucleus \citep{2009ApJ...703.2171S,2011MNRAS.415.1895E}. This AGN has shown moderately broad as well as narrow Fe-K lines near 6 keV \citep{2009ApJ...703.2171S}. The broadband X-ray spectrum including the soft X-ray excess and iron lines is well described by reflection from the accretion disc (see e.g., \citealt{2012ApJ...758...67L}). More recently, \citet{2016arXiv160205589L} have studied this AGN in detail using \xmm{} and \nustar{}~observations. They suggested that the soft excess emission may be explained by combining a blurred ionized reflection component and a spatially distinct Comptonization component.
 
 Fairall~9 has also been observed in  a number of multi-wavelength campaigns to probe the central engine including the broad line region  \citep{1985ApJ...297..151C,1985ApJ...295L..33W,1986ApJ...307..486M,1989ApJ...345..637K,1997A&AS..121..461R,1995ApJ...454L..11Z,
2012ApJ...749L..31L,2014ApJ...788...10L}. Some of the campaigns cover only  UV \citep{1985ApJ...297..151C,1997ApJS..110....9R,1989ApJ...345..637K} or UV and optical \citep{1985ApJ...295L..33W,1992A&A...256...33L,1997ApJS..112..271S} or UV and X-rays \citep{1995ApJ...454L..11Z,1986ApJ...307..486M} to search for possible correlation between continua at different wavelengths, and between the continuum and broad emission lines. Only a few multi-wavelength observations of this AGN have been reported to date to study the broadband from the optical to X-rays including the UV emission \citep{1997A&AS..121..461R,2014ApJ...788...10L}. \citet{2014ApJ...788...10L} studied nearly a 2.5-months of data from a \swift{} monitoring campaign which revealed the UV signature of X-ray flares. A detailed study of long monitoring campaign of such a``bare nucleus'' is lacking. Such a study will allow to investigate the different regions of the accretion disc. This AGN has been observed from time to time by \swift{} since its launch. 

In this paper, we use publicly available \swift{} monitoring observations performed over nearly two year period in the optical, UV and X-ray bands. We organize the paper as follows. We describe the observation and the data reduction in the next section. In section 3, we analyze the lightcurves and study cross-correlation between UV/optical and X-ray bands. We summarize our findings and discuss these results in section 4.

\section{Observations and data reduction}
We analyzed $\sim$ two years span of \swift{}~XRT \citep{2005SSRv..120..165B} and UVOT \citep{2005SSRv..120...95R} archival data. Both the XRT and UVOT instruments have simultaneous coverage in the X-ray and the UV/optical bands, respectively. We analyzed \swift{}~XRT data using the publicly available “Swift-XRT data product generator” provided on the University of Leicester website \footnote{http://www.swift.ac.uk/user\_objects/}. We adopted the methodology which is described in \citet{2007A&A...469..379E,2009MNRAS.397.1177E}. We used all observations from April 16, 2013 to April 23, 2015 excluding the 2008 observation. We extracted each count rate with the bin--size equal to the total exposure of a particular observation. The exposure of each observation is about $1ks$ (see in Table~\ref{obs_log}). We found usable 14 data points from photon counting (PC) mode and 133 data points in window timing (WT) mode excluding highly piled-up PC observations and ten snapshots with the improper centroid of the source in WT mode. Thus we found usable 147 pointings for XRT instrument.

\begin{figure} %\centering
\includegraphics[scale=0.85]{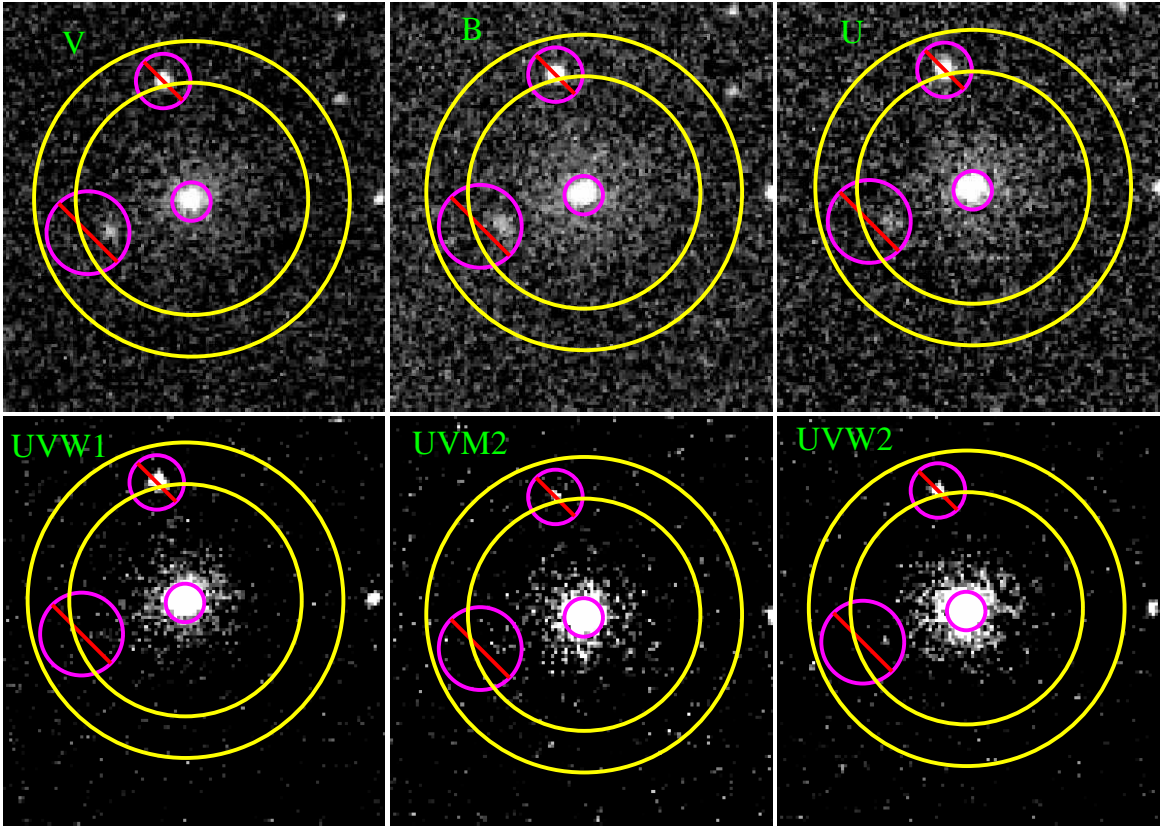}
\caption{00091908001: The selection of source and background regions from UVOT instrument. The selection in V to U bands in upper panel from left to right, and for UVW1 to UVW2 in lower panel from left to right are shown here. The crossed circular regions represent the two background sources which are excluded to estimate proper background contribution.}
\label{obs_net1} \end{figure}

We also analyzed UVOT \citep{2005SSRv..120...95R}~data sets acquired simultaneously with XRT. We found a few observations where XRT was not operated while UVOT was observing the source (e.g, 00091908031). Similarly, XRT was observing the source while UVOT was not used (e.g., 00037595053). We used sky image files to extract the count rate for an individual filter of a particular observation. The sky images are already corrected for any shift or rotation from the source position (RA $=20.940618$ degrees and DEC $=-58.805736$ degrees). We used {\tt UVOTIMSUM} to increase the signal to noise by summing multiple frame exposures. To extract the spectrum of each observation, we selected a circular source region of 6 arcsec centred at the source RA and Dec. We also selected the background region from an annular ring with inner radius 36 arcsec and outer radius 49 arcsec keeping same source centre in each frame exposure avoiding the two prominent UV/optical object as shown in Fig.~\ref{obs_net1}. We then used {\tt UVOT2PHA} which utilizes the {\tt UVOTSOURCE} task to compute fully corrected counts and area using latest calibration files \footnote{https://heasarc.gsfc.nasa.gov/ftools/caldb/help/uvot2pha.html}. We thus obtained spectrum products for both source and background, then extracted a background corrected source count rate and associated one sigma error for each passband of a particular observation. We found usable 165 pointings for UVOT instrument.

\begin{table*}
  \noindent \begin{centering}
    \caption{Observation log of \swift{}~XRT} \label{obs_log}
    \begin{tabular*}{15.0 cm}{lccccccc}
      \hline
      \hline 
       Observation-ID      & Date                                  & MJD--56000     & No. of IDs & On-target  & Epoch        \tabularnewline
                           &                                       &                &    (for XRT)        & XRT expo. (ks) &                 \tabularnewline      \hline
        00037595002--21    & April 16, 2013-- July 6, 2013         &398.7--479.9    &20          &23.7&  I   \tabularnewline      
        00037595022--50    & November 10, 2013-- January 1, 2014 &610.8--664.3    &28          & 27.9   &  II  \tabularnewline
       % 00037595041--50    & December 22, 2013-- January 1, 2014   &648.6--664.3    &10          & 9.0  & IIb          \tabularnewline
        00091908001--98    & April 2, 2014-- October 21, 2014     &749.0--951.9 &90              & 94 &III       \tabularnewline
        00037595052--70    & February 8, 2015-- April 23, 2015     &1061.8--1135.3 &9             & 3.5  &IV        \tabularnewline
      \hline
    \end{tabular*}
    \par\end{centering} {Note--UVOT observes all optical/UV filters except IVth epoch where only UVW2 filter was used.}
\end{table*}

\section{Data analysis}

\subsection{ X-ray and UV/optical lightcurves}
\begin{figure} %\centering
\includegraphics[scale=0.55]{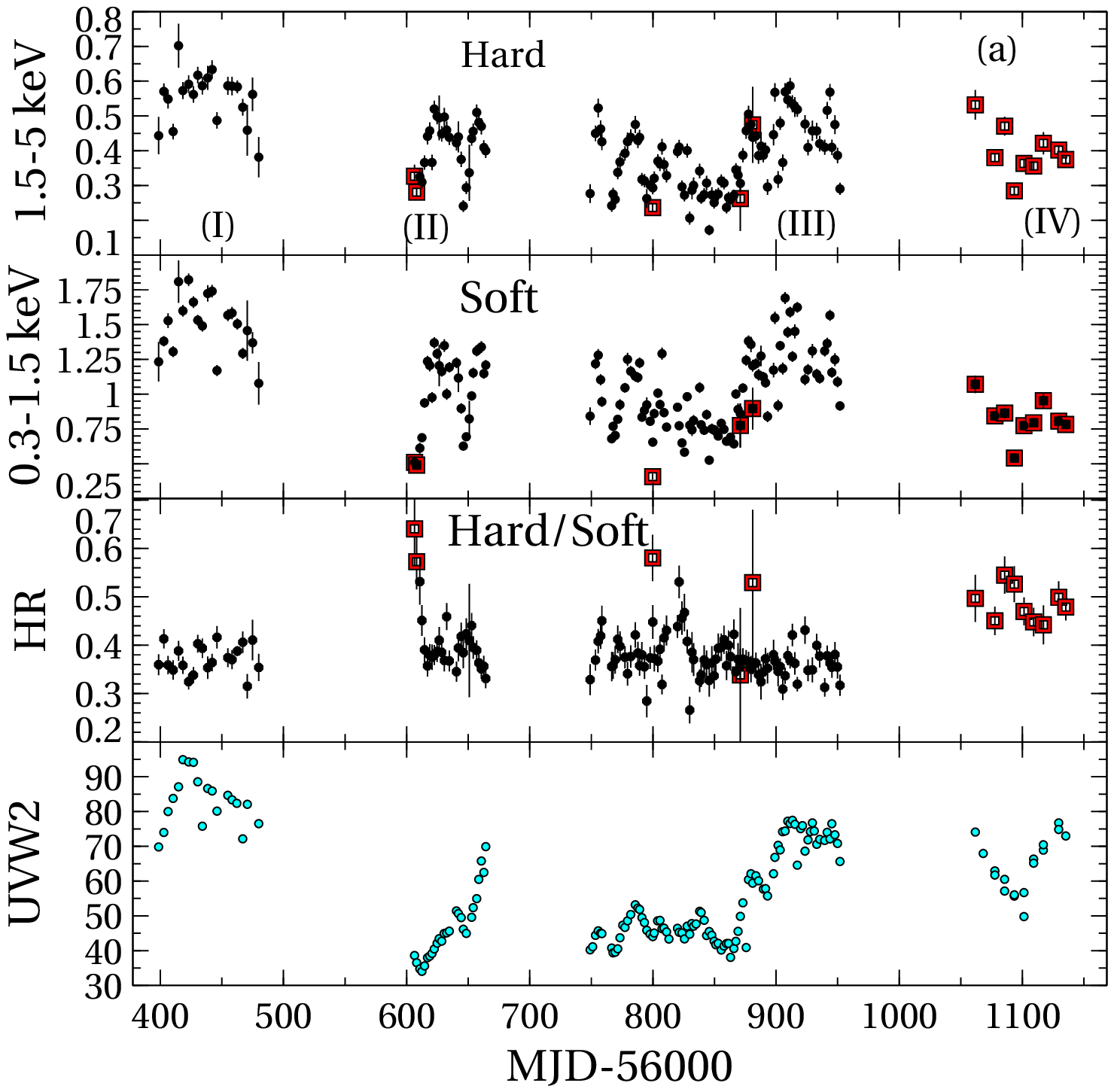}
\includegraphics[scale=0.55]{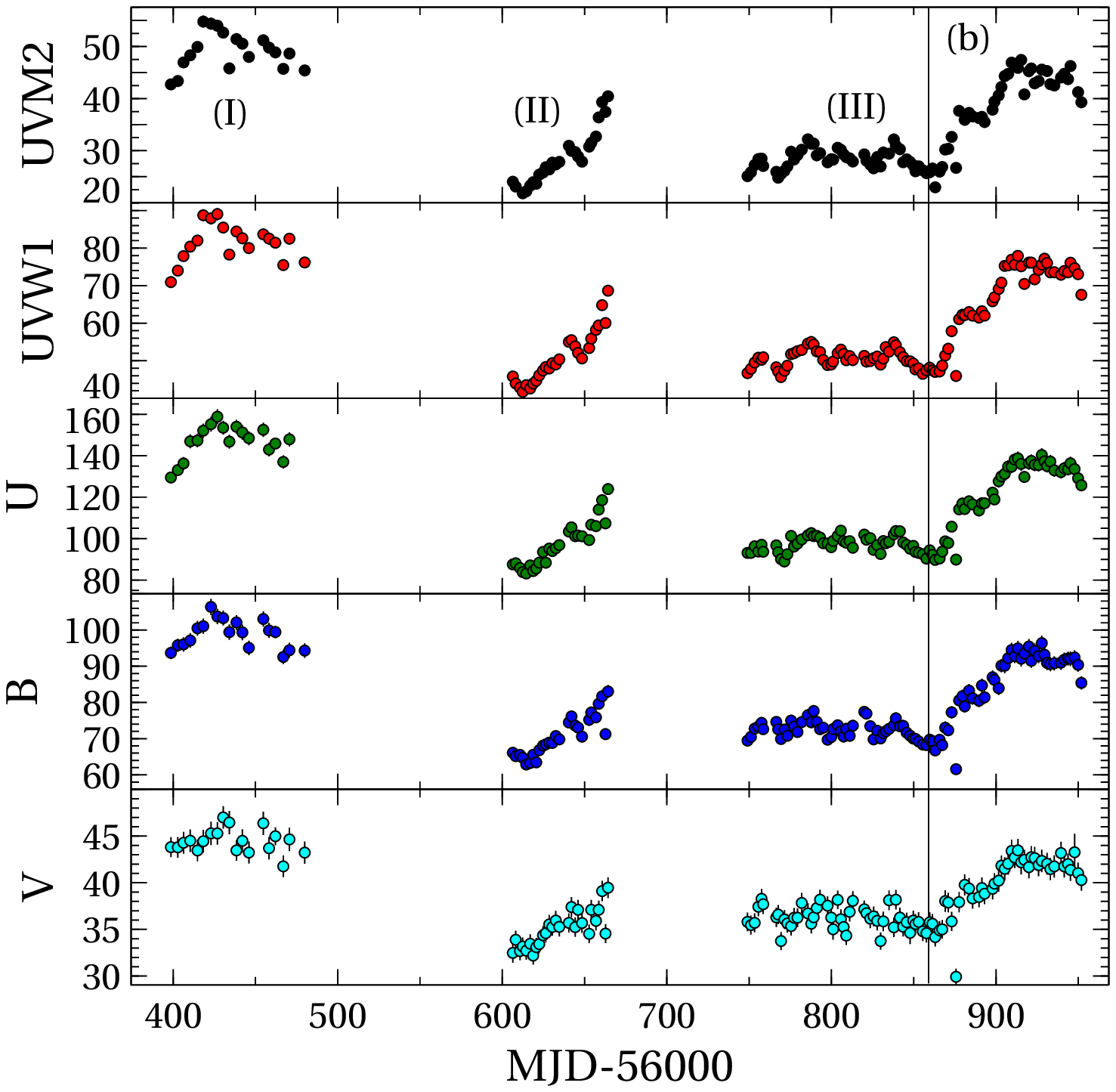}
\caption{ The count rates ($\rm counts~s^{-1}$) and hardness ratio for four time series which are marked by I, II, III and IV. (a) All X-ray count rates, corrected from systematic effects, observed by \swift{}/XRT~in the $1.5-5$\kev~(top panel) and $0.3-1.5$\kev~(second panel from top) bands. Their hardness ratio (third panel from top) and UVW2 count rates are displayed in the bottom panel. The data points with square marker correspond to photon counting (PC) mode and the rest of X-ray data points are observed in window timing (WT) mode. In IVth time series, the UVW2 data points are larger in number compared to PC mode data points as highly piled-up PC mode observations are excluded from the analysis. (b) In addition to UVW2, the count rates of the UV/optical ($\rm counts~s^{-1}$) observations simultaneous to X-rays are presented in UVM2, UVW1, U, B and V bands. The vertical lines mark feature such as rising trend in each lightcurve.}
\label{sw_obs} 
\end{figure}

We created lightcurves for the $1.5-5$\kev~ and $0.3-1.5$\kev~bands using their observed count rates in units of $\rm counts~s^{-1}$. We refer to the $1.5-5$\kev~as the hard X-ray band where simple power-law emission is considered to be the dominant component. We are limited to this energy band to mitigate the effect of reprocessed emission such as Fe-K$\alpha$ near 6\kev~detected by, for example, \citet{2009ApJ...703.2171S}. The $0.3-1.5$\kev~band is referred to as the soft X-ray band which may consists of with multiple complex spectral features due to warm absorption, blurred reflection and possible intrinsic disc Comptonization. All the observed epochs are marked by I, II, III and IV for different epochs which form the full lightcurve for Fairall~9. For simplicity, the time on the x-axis is shown in MJD after subtracting by 56000. All four epochs have been observed simultaneously in the X-ray to UV/optical bands. Each observation of epoch I was taken on a cadence of four days. The epochs II and III were carried out at the best sampling to date on two days cadence while epoch IV consists of the observations acquired on weekly basis. In Fig.~\ref{sw_obs}~(a), the $1.5-5$\kev,~ $0.3-1.5$\kev~lightcurves and their hardness ratio (1.5-5 keV/0.3-1.5 keV) are shown for all observed epochs to date. The X-ray emission varies by a factor of $\sim 4$ from minimum to maximum on long timescale and also varies by a factor of $\sim2$ from minimum to maximum on short timescale. The hardness ratio seems constant on long timescales while it appears to vary rapidly on short timescales.  

 The UVW2 emission has simultaneous coverage similar to the X-ray emission as shown in the bottom panel of Fig.~\ref{sw_obs}~(a). The UVW2 observations for epoch IV are larger in number as highly piled-up photon counting mode data sets for the X-ray emission are excluded from the analysis. The other filter (UVM2, UVW1, U, B and V) observations only cover simultaneously to the X-rays in epochs I, II and III which are shown in Fig.~\ref{sw_obs}~(b). We estimated the variations on long timescales relative to the observed minimum value for each individual epoch and full epoch (I+II+III+IV), and we also estimated the maximum variation on short timescales for individual epoch as listed quantitatively in Table~\ref{values}. Thus all the epochs are highly variable on both long and short timescales. In addition to rapid variability on timescales of few days to a week, there is a slowly rising emission component present in all the observed lightcurves. This is clearly visible in each lightcurve in the 850--950 days of MJD-56000.

\begin{table*}
\caption{Long and short--term variability for all observed lighcurves.} \label{values}

\begin{tabular}{cccccccccccc}
\hline 
  & \multicolumn{8}{c}{\% variability on long~(L) \& short (S)~timescales} \tabularnewline
\hline 
\hline 
Bands &\multicolumn{2}{c}{I}&\multicolumn{2}{c}{II} &\multicolumn{2}{c}{ III}&\multicolumn{2}{c}{IV} & \multicolumn{2}{c}{Full}\tabularnewline
\hline
              & L  &S    &L&S    &L&S    &L&S    &L&S \tabularnewline
 %     &  (a) & (b)              & \tabularnewline
\hline 
$1.5-5$ keV   & 184.1&54.3&215.8&42.0&341.4&78.9&187&64.4&408.7&--\tabularnewline
$0.3-1.5$ keV & 169.1 &48.8 &279.3&40.3 &415.3&68.7&198.1&59.8&371.7&--\tabularnewline 
UVW2            &136.2&16.9 &205.6 &11.9 &203.8&48.0&54.1&33.2&100&--\tabularnewline
UVM2            &128.4&15.0 &185.0 &9.8 &206.8&41.2&--&--&251.2&--\tabularnewline
UVW1           &126.2 &9.7 &164.6 &14.3  &172.0&33.2&--&--&213.5&--\tabularnewline
U            & 127.1& 9.1& 149.1& 15.4&164.6&27.5&--&--&190.9&--\tabularnewline 
B            & 125.1&7.3 &132.7 &17.0 &159.8&33&--&--&172.8&--\tabularnewline 
V            & 112.0&7.8 &122.9 &14.8  &154.2&34.1&--&--&157.1&--\tabularnewline
\hline 
\end{tabular}	
\end{table*}

%\begin{landscape}
%\end{landscape}

% \subsection{Pearson Product Moment Correlation coefficient ($r$)}
In order to study the linear correlation between different wavelengths due to rapid variations and common trends present in each observed lightcurve, we used Pearson's correlation coefficient~`$r$'. It is also called as  Pearson Product Moment Correlation coefficient (PPMC). The measurement of coefficient `r' is the strength of correlation between two variables. The significance of the correlation can not be measured alone by estimating the correlation coefficient `r'. If we assume each variable from $n$ data points associated with each respective data set follows a normal distribution, the t-statistic with $n-2$ degree of freedom is used to provide the significance of the correlation. The t-statistic for a given correlation coefficient `r' is defined as $t^2=(n-2)~\frac{r^{2}}{1-r^2}$. This t-statistic allows us to determine the probability that the correlation occurs by chance and the probability is termed as $p$ value. Small values of $p$ (i.e., $<0.05$) indicate that the observed variables are likely correlated. We determined the correlation coefficient between different wavebands and the probability
using the Python routine for PPMC. We found coefficient $r$ in the range 0.92--0.63 for full lightcurve with log(p) values in the range -57.9-- -14.8, and for II+III time series, $r$ covers 0.91-0.43 with log(p) values in the range -45.4-- -5.5. There are therefore strong correlations between the X-ray emission as well as UV/optical emission.

\subsection{Z-transformed Discrete Cross-correlation function (ZDCF)}
\begin{figure} %\centering
\includegraphics[scale=0.7]{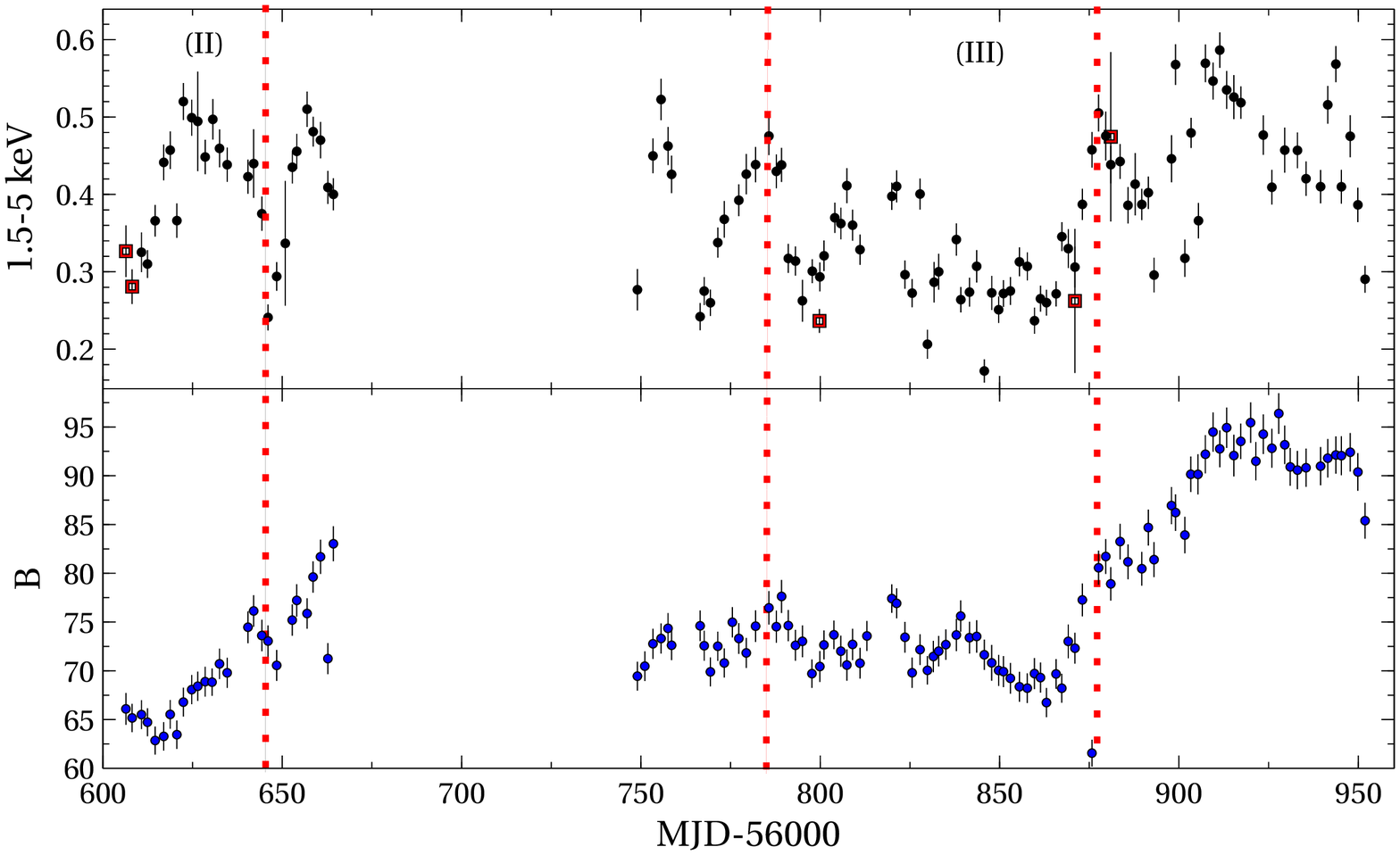}
\caption{The count rates ($\rm~counts~s^{-1}$) for II+III time series in the $1.5-5$\kev~(top panel) and the B band (bottom panel). Clearly, the emission in the B band seems shifted version of the X-ray emission. The vertical lines mark features such as observed shift in the lightcurves.} \label{iipiii} \end{figure}

The observed lightcurves in the UV/optical bands appear to be shifted compared to the X-ray lightcurves. The shift can be seen between B band and the hard X-ray band as shown by marked vertical lines in Fig.~\ref{iipiii}.  Various tools such as Z-transformed discrete correlation function \citep[ZDCF][]{1997ASSL..218..163A} and Discrete Cross-correlation function \citep[DCF:][]{1988ApJ...333..646E} are commonly used to estimate the shift or time delay between a pair of bands. The time series for astronomical data is normally found sparse and unevenly distributed. 
The ZDCF is used to estimate the cross-correlation of sparse, unevenly sampled lightcurves . This method improves several weaknesses of DCF.  The ZDCF uses equal population binning and Fisher's z-transform. If there are $n$ pairs in a time-lag bin, the CCF ($\tau$) is computed by the correlation coefficient

\begin{equation}
r_z=\frac{\sum_{i}^{n}(a_{i}-\underline{a})(b_{i}-\underline{b})/(n-1)}{s_{a}s_{b}}%\,.\label{e:rbin}
\end{equation}

 where $\underline{a}$, $\underline{b}$ are the estimators of the bin averages, and
$s_{a}$, $s_{b}$ the estimators of the standard deviations which are difined as

\begin{equation}
s_{a}^{2}=\frac{1}{n-1}\sum_{i}^{n}(a_{i}-\underline{a})^{2}\
\end{equation}

Using Fisher's transformation which provides the normal distribution of $z$--values for given $r_z$ values.

\begin{equation}
z=\frac{1}{2}\log\left(\frac{1+r_z}{1-r_z}\right)%\,,\,\,\,\zeta=\frac{1}{2}\log\left(\frac{1+\rho}{1-\rho}\right)\,,\,\,\, r=\tanh z\,,\label{e:z}
\end{equation}

After estimating the mean ($\underline{z}$) and standard deviation ($s_{z}$) of $z$, the transforming it back to calculate  $\pm1\sigma$ error in $r_z$ is approximately equal to

\begin{equation}
dr_{z\pm}=|\tanh(\underline{z}(r_z)\pm s_{z}(r_z))-r_z|\,.\label{e:dcferr}
\end{equation}

\begin{figure} \centering
\includegraphics[scale=1]{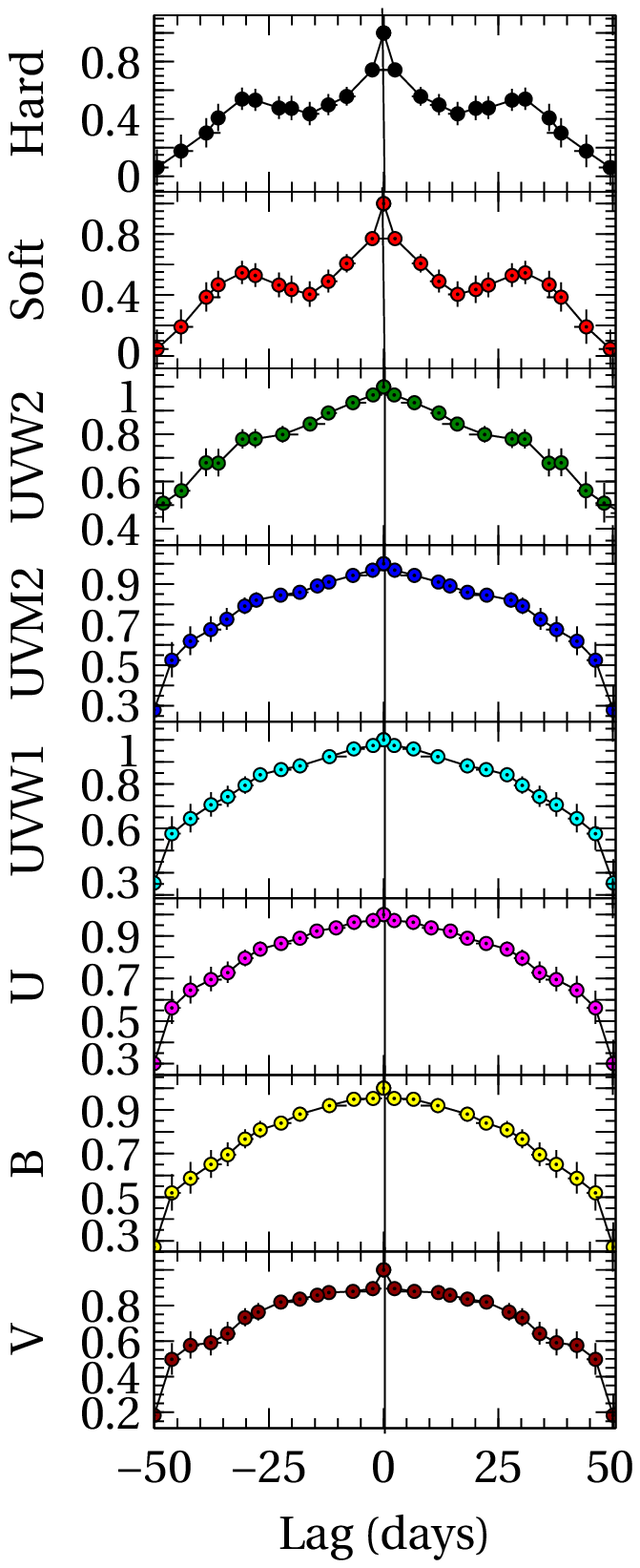}
\includegraphics[scale=1.04]{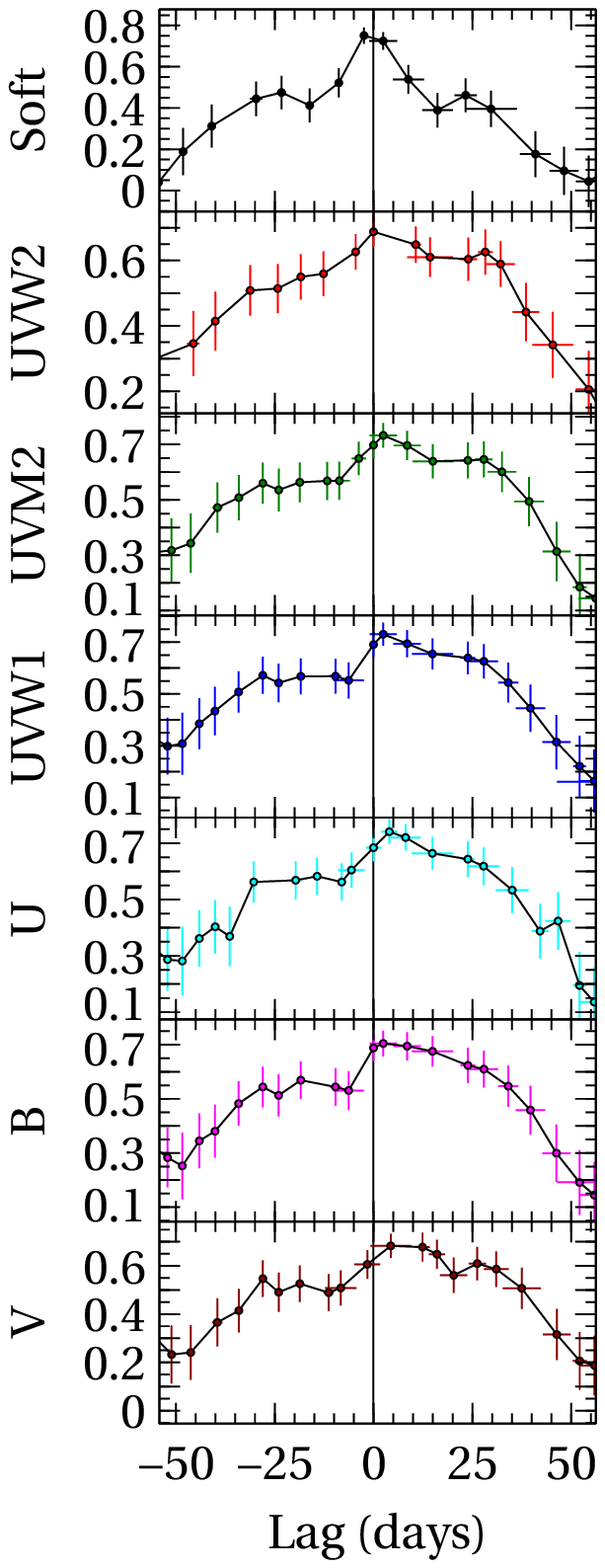}
\caption{Left panel: Auto Correlation function (ACF) of hard to UV/optical lightcurves using Z-transformed Dicrete cross--correlation function. Right panel: the cross-correlation function (CCF) is shown for each UV/optical and soft X-ray bands with respect to hard X-ray band ( $1.5-5$\kev). The CCF represent that the soft emissions from soft X-ray to V bands lag the hard X-ray. The vertical solid lines represent zero delay in all graphs.}
\label{zdcf}
\end{figure}

The ZDCF method has advantage over DCF as this can estimate the error based on Gaussianity of data points. In general, the ZDCF tool uses a variable bin size to ensure at least 11 data points per bin. In this work, we made use of minimum 50 data points together in a time bin and then produced 10000 realizations through Markov Chain Monte Carlo (MCMC) to estimate the lag for a particular soft band with respect to the hard band. In addition, we also used non zero-lag data points and non-uniform binning to compute the cross-correlation function (CCF), and for auto cross-correlation function (ACF) we used all data points plus non-uniform binning. The results of ACF and CCF are shown in the left and right panels of Fig.~\ref{zdcf}, respectively, in the range (-50, 50) days. Interestingly, ACF of each UV/optical bands is much broader than that of the X-ray bands and the CCF becomes more broader with wavelength on positive delay. Clearly, the peak in each CCF seems to shift in positive side of the curves. We derived the lags about the peak in the range -100 to 100 days for all time series together (full) and II+III using "PLIKE~\footnote{http://wwo.weizmann.ac.il/weizsites/tal/research/software/}" code provided by \citet{1997ASSL..218..163A}. We found the lags in days for each band from UV/optical to soft X-ray relative to the hard X-rays as listed in Table~\ref{lags}.   

\subsection{{\tt JAVELIN}: Lag estimation}
%\subsubsection{lightcurve modeling}
We used the model-independent tool like ZDCF to estimate the lag between various energy bands. The lag is important to understand the system physically if one can estimate the lag using the model-- dependent method such as damped random walk (DRW) process. The DRW method introduced by \citet{2009ApJ...698..895K} has been used to derive the quasar UV/optical variability since last few years  \citep{2010ApJ...708..927K,2010ApJ...721.1014M,2011ApJ...735...80Z, 2013ApJ...765..106Z}. This is a stochastic process, defined by an exponential covariance matrix ($C_{ij}$) that works as a random walk for a short timescale and asymptotically obtains a finite variability amplitude on long timescale \citep{2013ApJ...765..106Z}. 

\begin{equation}
C_{ij}(\Delta t_{ij})=\sigma^{2}_{d}~ e^{-\frac { \left| {\Delta t_{ij}}\right|}{\tau}}
\end{equation}
Where ${\Delta t_{ij}}=t_{i}-t_{j}$, and $\sigma_{d}$ and $\tau$ are model parameters. The covariance is $\sigma^{2}_{d}$ for $\Delta t_{ij}\ll \tau$, and decays exponentially on a timescale given by $\tau$. 
 A publicly available Python code {\tt JAVELIN} assumes one lightcurve is a DRW process and the second is a convolution of the DRW lightcurve with a uniform delay distribution sepcified by mean delay $\tau$ and delay range $\Delta\tau$. It has been seen that DRW and X-ray power spectral slopes for AGN are similar on short timescales (i.e.,\citealt{ 2005MNRAS.359.1469M,2004MNRAS.348..783M}). We therefore used  {\tt JAVELIN} to derive the probability distribution of lag for various wavelengths. This code calculates a maximum likelihood lag, scale factor and a kernel width assuming a top-hat transfer function from the DRW covariance matrix. This method also employs internally a linear detrending procedure to remove the mean of the lightcurve used, and hence further additional substraction is not required. 

\begin{figure} %\centering
\includegraphics[scale=0.35]{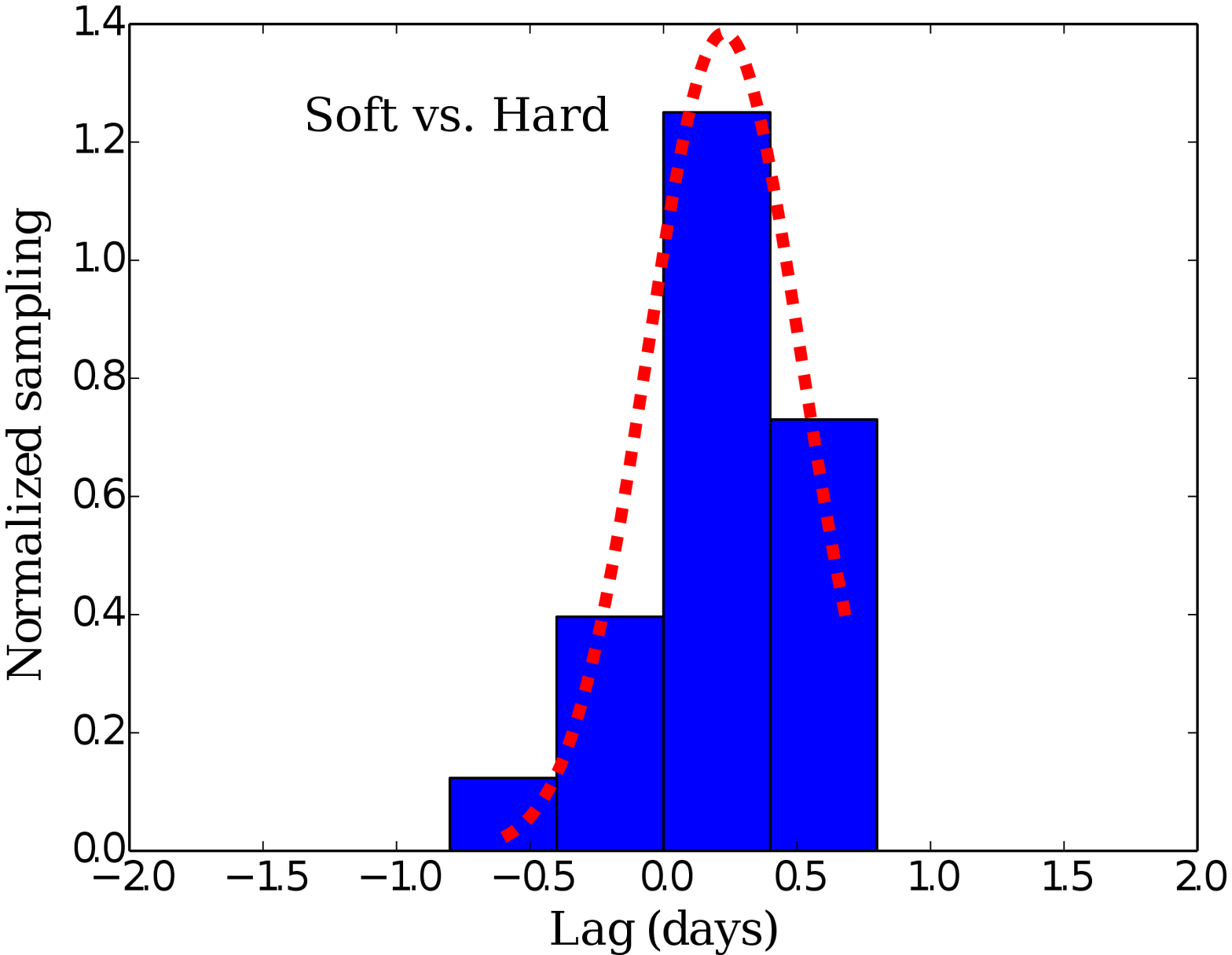}
\includegraphics[scale=0.35]{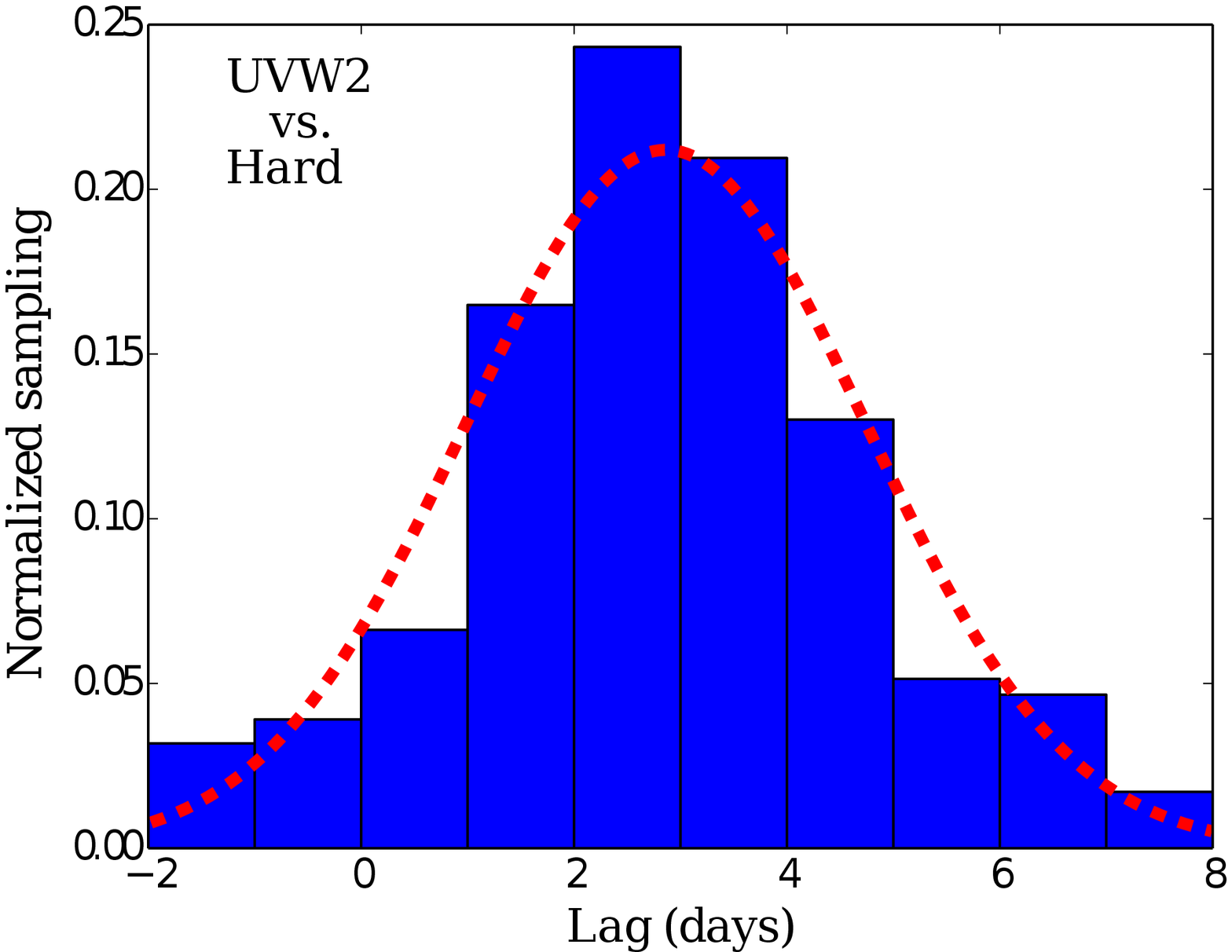}
\includegraphics[scale=0.35]{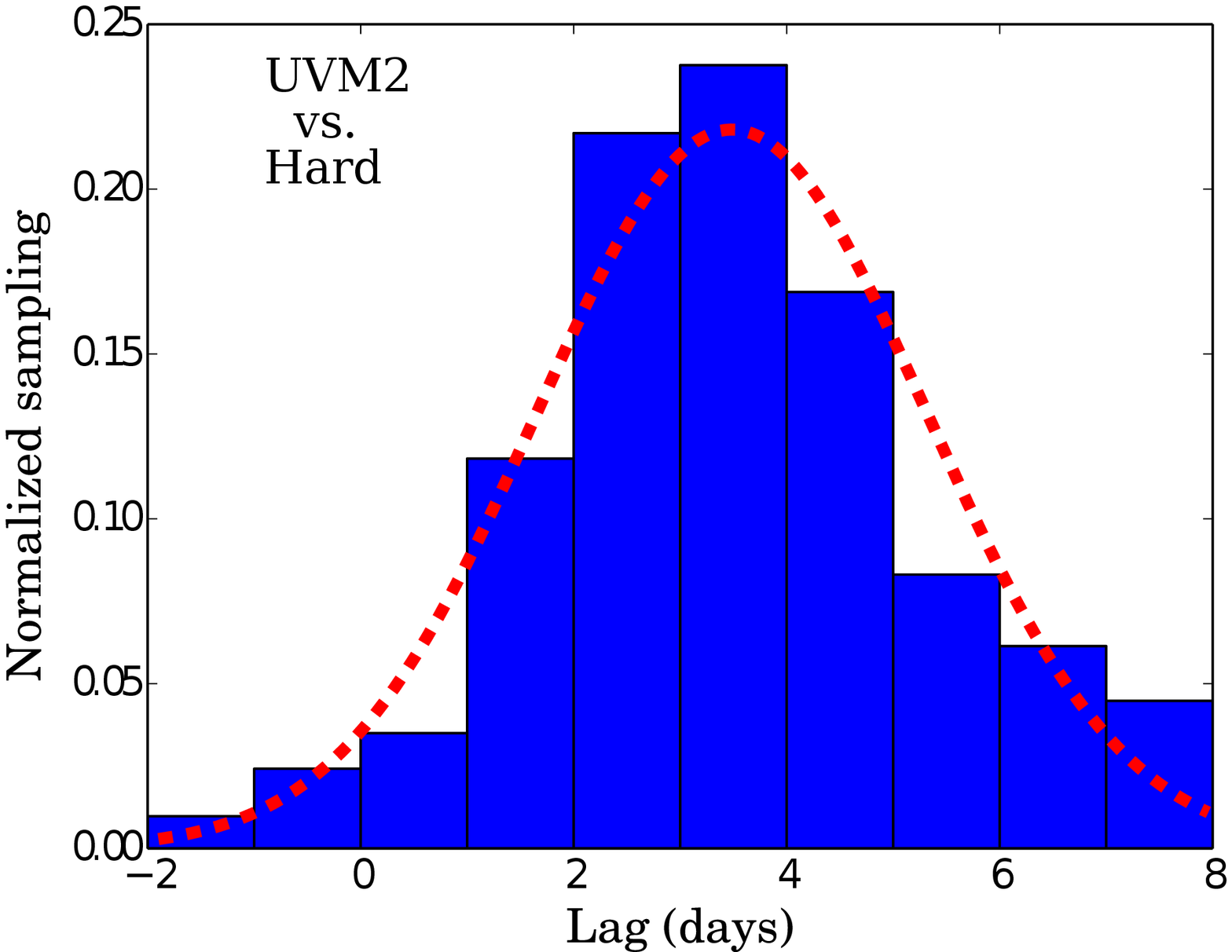}
\includegraphics[scale=0.35]{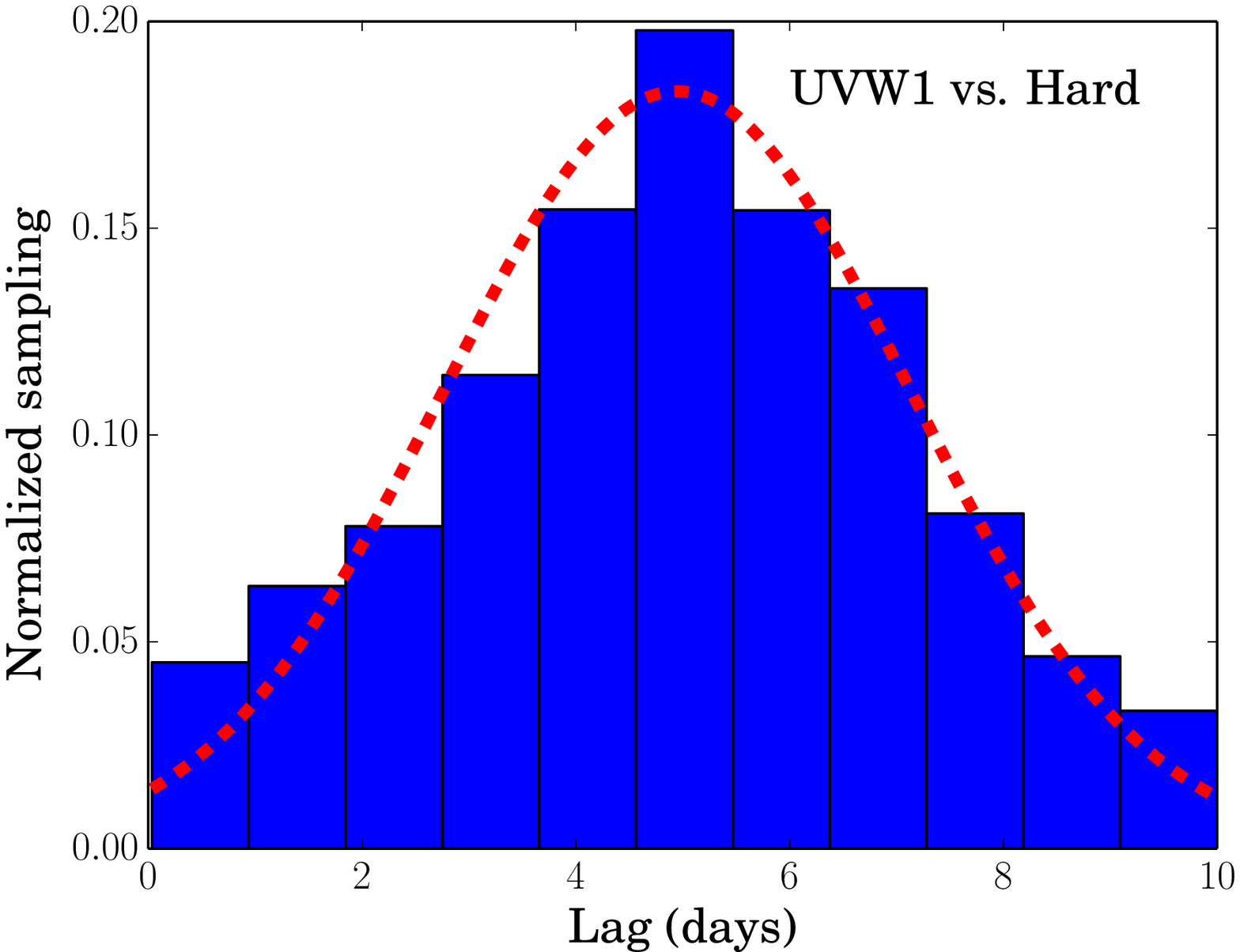}
\includegraphics[scale=0.35]{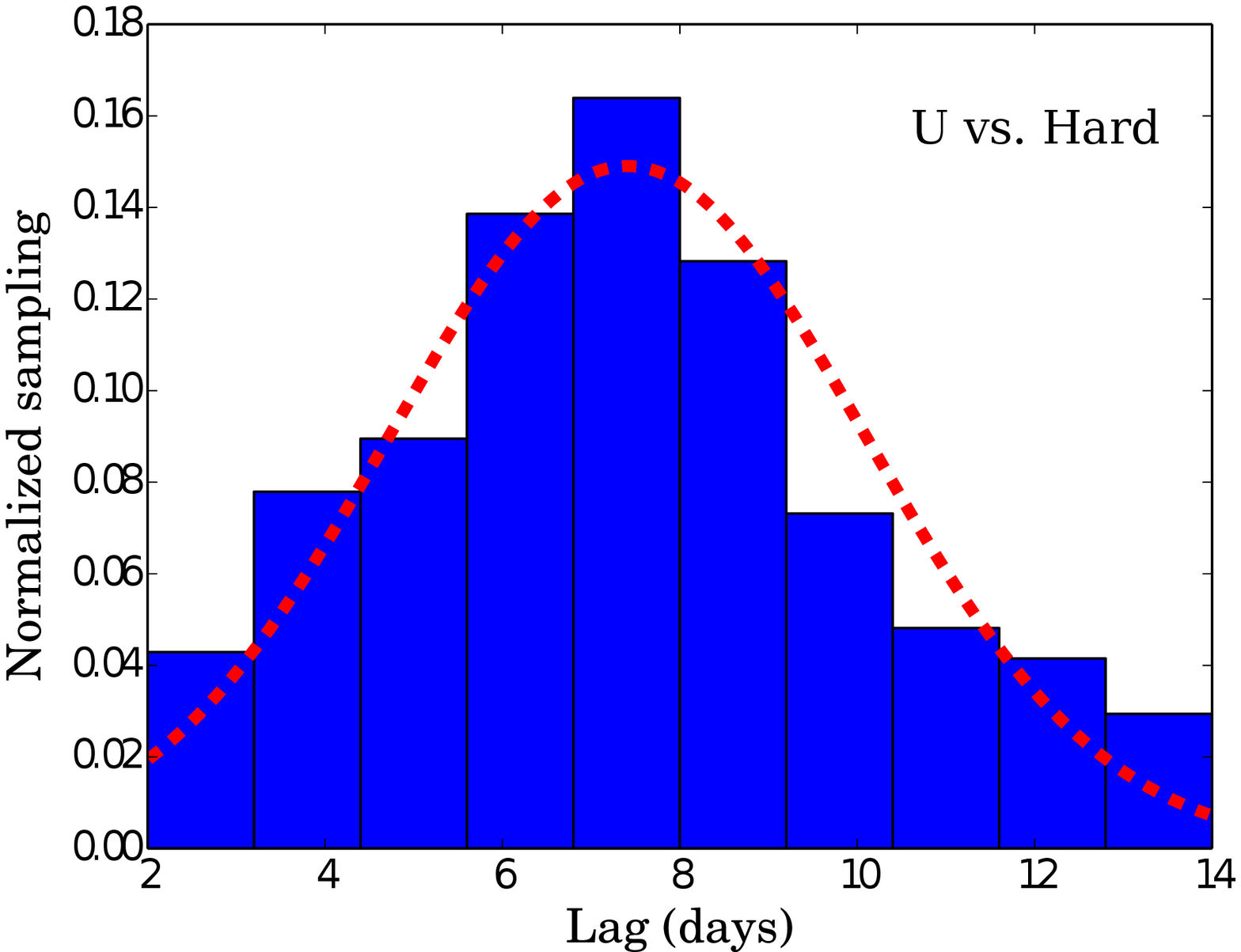}
\includegraphics[scale=0.35]{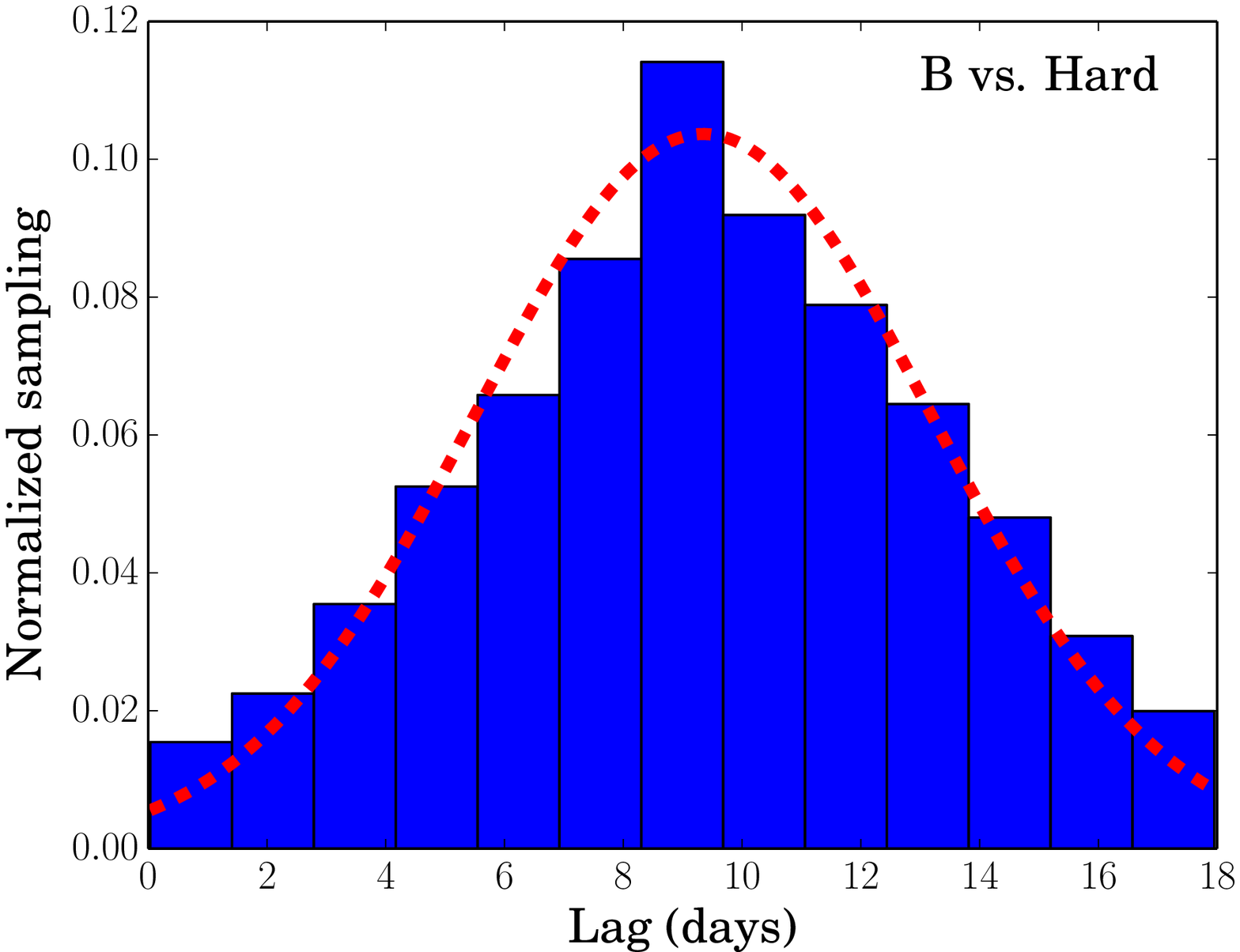}
\includegraphics[scale=0.35]{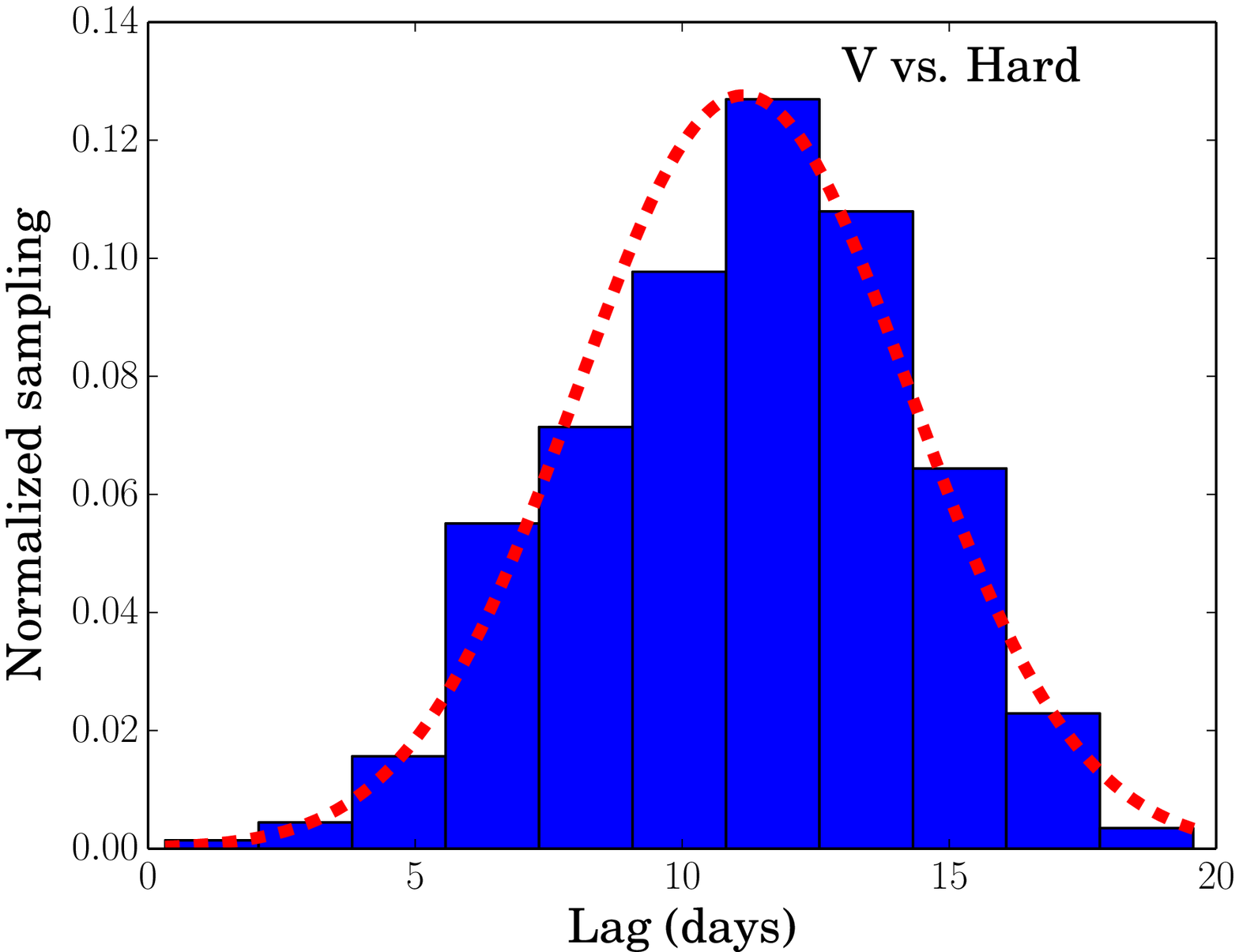}
\caption{a) {\tt JAVELIN}: probability distribution of lags from soft X-ray to V band emission with respect to the hard X-rays for full lightcurve including I+II+III+IV. The red curve represents the Gaussian representation of probability distribution of observed lag.}
\label{javelin}
\end{figure}

We used the hard X-ray ($1.5-5$ keV) as the first lightcurve to construct the continuum parameters ($\sigma^2_d$ and $\tau$) and the soft band (i.e., soft X-rays, UVW2, UVM2, UVW1, U, B, V) lightcurve as the second lightcurve to estimate delay using reverberation model (Rmap\_Model \footnote{https://bitbucket.org/nye17/javelin}). We also created 10000 realizations of lags between the first and second lightcurves using the MCMC method embedded in {\tt JAVELIN} code. The probability distribution of lags for full lightcurve (I+II+III+IV) is displayed in Fig.~\ref{javelin} derived with respect to the hard X-ray band. Assuming a Gaussian distribution of lags, we computed the mean and associated one $\sigma$ error for each pair of UV/optical and soft X-ray with the hard X-rays. Similar to ZDCF, we used both full and II+III epochs to derive the lags which are listed in Table~\ref{lags}. Thus we obtained a clear increasing trend of lags with wavelength.

\subsection{Lag spectrum and its modeling}
\subsubsection{Lag spectrum}
Since all the UV/optical bands are generlly represented by their central effective wavelength, we converted the soft X-ray band (0.3-1.5 \kev) and the hard X-rays ($1.5-5$ \kev) into average wavelengths equal to $1.4\pm2.1~nm$ and $0.4\pm0.7~nm$, respectively. The estimated lags by ZDCF and {\tt JAVELIN} for the UVW2, UVM2, UVW1, U, B, V and $0.3-1.5$\kev~band are consistent within errors. Here we used {\tt JAVELIN} lags obtained assuming a Gaussian distribution from the full time series. We created the lag spectrum as a function of wavelength measured in $\rm~nm$ after transforming the lags with respect to the UVW2 band. This is shown in Fig.~\ref{lag_spec} by black points with red error bars.

\subsubsection{ Power-law model and linear model}
 The functional form of commonly used thin disc model as the power-law model which is given by Equation~\ref{equ:lag_wv}, where $\tau$ is the time delay corresponding to a particular wavelength $\lambda$ with respect to a reference wavelength $\lambda_{0}$, $\alpha$ is the normalization and  and $\beta$ is the index. 
\begin{align}
\label{equ:lag_wv}
\tau = \alpha\left[\left(\frac{\lambda}{\lambda_0} \right)^{\beta}
  -1\right]
\end{align}
The value of $\beta$ is $4/3$ for standard accretion disc model. In our case we used the reference wavelength $\lambda_{0}=192~nm$ for the UVW2 band. We fitted the observed lag spectrum by power-law model (see Equation~\ref{equ:lag_wv}). The best--fitting model was found to be $\tau = 3.0\pm0.4\times \left[\left(\frac{\lambda}{192} \right)^{1.36\pm0.13}-1\right]$. We also modeled the observed lag spectrum with the linear function A+B$\times \lambda$. The linear model $-2.9\pm0.4+0.020\pm0.001\times\lambda$ was found to fit the lag spectrum equally well. However, both power-law and linear model provide similar poor statistics ($\chi^{2}_\nu/dof=0.1/4$) to describe the observed lag spectrum. Both the fitted power-law model and linear model are shown by black and blue curves, respectively, in Fig.~\ref{lag_spec}.

\subsubsection{Standard accretion disc model}

 The UV/optical emission can be considered as the mixture of both the emission from the standard disc itself and the themalized emission in the disc due to X-ray heating. The thermalized emission in the accretion disc increases temperature of the accretion disc. As a result, the increased temperature can be observed in the UV/optical bands. In the X-ray heating, we suppose a compact X-ray source at a height $H$ above the accretion disc close to the SMBH. We also assume the X-ray luminosity $L_{X}$ and albedo $A$ for X-ray heating in the disc. 
 The resultant temperature ($T$) due to the viscous heating as well as the X-ray heating at disc radius $R$ from the centre can be described as given below in Equation~\ref{Tprofile1} (see e.g., \citealt{2007MNRAS.380..669C,2000ApJ...535..712B}).
\begin{align}
\label{Tprofile1}
T(R) = \left(\frac{3GM\dot M}{8\pi \sigma R^3} (1-\frac{R_{in}}{R})^{1/2}+ \frac{(1-A)L_{\rm X}}{4\pi
    \sigma R_{X}^2} ~\rm cos~(\theta_{X})\right)^{1/4}
%\label{TR}
\end{align}
where $G$ is the gravitational constant, $M$ is the mass of the black hole, $\dot M$ is the accretion rate, $R_{in}$ is the innermost circular radius and $R_{X}$ is the distance of the disc element from the X-ray source. The parameter $\theta_{X}$ is the angle between the line joining of the disc element and the X-ray source, and the normal to the disc surface. The terms $cos (\theta_{X})$ and $R_{X}$ can be written as $\frac{H}{R_{X}}$ and  $(H^2+R^2)^{1/2}$, respectively.

After replacing parameters $cos (\theta_{X})$ and $R_{X}$ in Equation~\ref{Tprofile1} in terms of $H$ and $R$, Equation~\ref{Tprofile1} becomes as the following.
\begin{align}
\label{Tprofile2}
T(R) = \left(\frac{3GM\dot M}{8\pi \sigma R^3} (1-\frac{R_{in}}{R})^{1/2} + \frac{(1-A)L_{\rm X} H }{4\pi
    \sigma (H^2+R^2)^{3/2}} \right)^{1/4}
%\label{TR}
\end{align}
When the height of the X-ray source and innermost circular radius are much smaller than $R$, $(H^2+R^2)^{1/2}\sim~R$ and $(1-\frac{R_{in}}{R})^{1/2} \sim1$, the temperature ($T$) can be described as given in Equation~\ref{Tprofile} . 

\begin{align}
\label{Tprofile}
T(R) = \left(\frac{3GM\dot M}{8\pi \sigma R^3} + \frac{(1-A)L_{\rm X} H }{4\pi
    \sigma R^3} \right)^{1/4}
%\label{TR}
\end{align}

%Using height and radius in units of $r_g$ and accretion rate relative to the Eddington rate, the derived Equation~\ref{Tprofile} can be written similar to the equation 6.27 given in Cameron's Ph.D. thesis\footnote{http://eprints.soton.ac.uk/366670/1.hasCoversheetVersion/Cameron.pdf} when height and innermost radius are much smaller compared to the distance $R$. 

We write the radius $R$ in terms of temperature, mass ($M$), accretion rate ($\dot{M}$), height ($H$), albedo ($A$) and the X-ray luminosity ($L_{X}$). The delay is expressed by Equation~\ref{tauR} after converting temperature into wavelength from Wein's law, and $R$~as a product of light speed $c$ and delay $\tau$.   
\begin{align}\label{tauR}
  c\tau =
  \left(\frac{\lambda}{k}\right)^{4/3}\left(\frac{3GM\dot
      M}{8\pi \sigma } + \frac{(1-A)L_{\rm X} H }{4\pi \sigma} \right)^{1/3}.
\end{align}
where $k=2.9\times10^{-3}~mK$. Considering UVW2 wavelength as the reference wavelength $\lambda_{0}=192.8 \rm~nm$, we write the lags as a function of wavelength (see Equation~\ref{equ:MmdotLx}).       
\begin{align}
\label{equ:MmdotLx}
\tau - \tau_0 =
\left(\frac{1}{c}\right)\left(\frac{\lambda_0}{k}\right)^{4/3}\left(\frac{3GM\dot M}{8\pi \sigma } + \frac{(1-A)L_{\rm X} H }{4\pi \sigma}  \right)^{1/3}\nonumber \\
\left[\left( \frac{\lambda}{\lambda_0} \right)^{4/3} - 1  \right].
\end{align}

\begin{figure}
\centering
\includegraphics[scale=0.6]{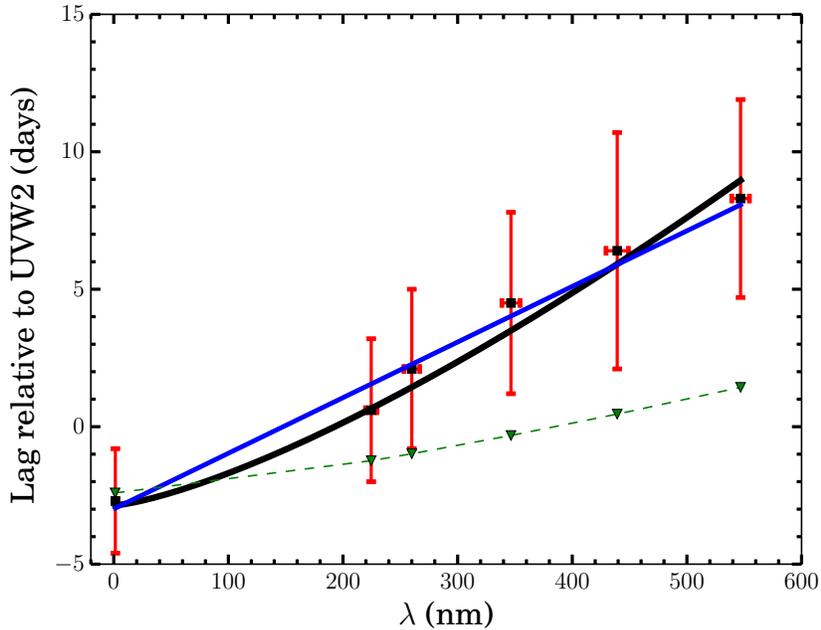}
\caption{The lag spectrum is created after transforming the observed lags for all energy bands with respect to the observed lag of UVW2 band. The fit for the lag spectrum (a) Black curve- the best-fit power-law model ($2.9\pm0.4\times[(\lambda/192.8)^{1.36\pm0.13}-1]$) (b) Blue line: the best-fit linear model ($-3.0\pm0.4+0.020\pm0.001\times\lambda$) (c) Green dashed curve : standard accretion thin disc theoretical model for Fairall~9.}
\label{lag_spec}
\end{figure}

 We estimated the predicted lags from the standard  disc by using Equation~\ref{equ:MmdotLx}. We used currently available mass of black hole $M_{BH}=2.6\times10^8~M_{\odot}$, height $H=6r_{g}$ and inner radius $R_{in}=6r_{g}$ and albedo $A=0.2$. We also used the accretion rate relative to Eddington rate $\dot{m_{E}}=0.02$ and luminosity $L_{x}=10^{44.8}$\ergsec~ \citep{2009MNRAS.392.1124V}. Since we transformed the lags with respect to the reference wavelength $\lambda_{0}$, then we used $\tau_{0}=0$ days in the theoretical calculation. The theoretical calculation is shown by green dashed line and solid triangle corresponding to the central effective wavelength on the curve in Fig.~\ref{lag_spec}. This calculation reveals that the observed lags appear longer than the expected lags from the standard accretion disc.

Furthermore, Equation~\ref{equ:MmdotLx} can provide the radius of the annulus emitting region of an accretion disc for a given reference wavelength if we assume that the accretion disc is face on (e.g., \citealt{2015ApJ...806..129E}). After comparing Equations ~\ref{equ:lag_wv}~and ~\ref{equ:MmdotLx} when $\tau_{0}=0$ for the reference wavelength $\lambda_{0}=192\rm~nm$, we found that the normalization of Equation~\ref{equ:lag_wv} can be expressed as

\begin{align}
\alpha=
\left(\frac{1}{c}\right)\left(\frac{\lambda_0}{k}\right)^{4/3}\left(\frac{3GM\dot M}{8\pi \sigma } + \frac{(1-A)L_{\rm X} H }{4\pi \sigma}  \right)^{1/3}.
\label{equ:alpha}
\end{align}

The best--fitting of power-law model has a value of $\alpha\sim3.0$ days. The radius of annulus region over accretion disc may be about $\sim3.0\times86400\times3\times10^{10}\sim7.8\times10^{15}~cm$. The radius of annulus region of the disc can be described in terms of $r/R_{S}\sim400$ in the vicinity of the SMBH, where $R_{S}$ is Schwarzchild radius for Fairall~9 ($M_{BH}\sim2.6\times10^{8}~M_{\odot}$). The estimated temperature $T\sim1.5\times10^{3}~K$ corresponding $\lambda_{0}=192.8~nm$ using Wien's law peaks at above radius  of the accretion disc .

\begin{table*}
\caption{The observed lags for the soft bands estimated from ZDCF and JAVELIN tools with respect to the hard X-ray band.} \label{lags}
\begin{tabular*}{9.0 cm}{cccccc}
\hline 
&&\multicolumn{4}{c}{Full (I+II+III+IV) and II+III lightcurves} \tabularnewline
\hline 
\hline 
Bands &\multicolumn{2}{c}{ZDCF}&\multicolumn{2}{c}{JAVELIN} \tabularnewline \hline
   & Full& II+III   & Full  & II+III \tabularnewline
\hline 
0.3-1.5 \kev    &$1.4_{-1.9}^{+5.4}$ &$0.0\pm0.9$ & $0.23\pm0.29$&$-0.03\pm0.26$\tabularnewline
UVW2            &$1.3_{-4.0}^{+6.9}$ &$3.9_{-2.8}^{+6.2}$&$2.9\pm1.9$ &$4.4\pm1.5$\tabularnewline
UVM2             &$6.4_{-6.4}^{+15.4}$  &$2.5_{-2.2}^{+5.5}$&$3.5\pm1.8$&$3.8\pm1.9$\tabularnewline
UVW1             &$4.3_{-4.5}^{+8.1}$  &$3.9_{-3.0}^{+7.4}$&$5.0\pm2.2$&$6.7\pm2.6$\tabularnewline
U              &$2.4_{-3.1}^{+9.6}$  &$3.9_{-2.6}^{+8.0}$&$7.4\pm2.7$&$11.7\pm5.2$\tabularnewline 
B              &$4.3_{-5.1}^{+10.2}$ &$11.9_{-7.6}^{+13.2}$&$9.3\pm3.9$&$12.6\pm4.4$\tabularnewline 
V              &$4.0_{-2.7}^{+7.9}$ &$6.0_{-3.1}^{+9.2}$&$11.2\pm3.1$&$11.9\pm3.2$\tabularnewline
\hline 
\end{tabular*} 
\end{table*}

To summarize, we have estimated the extent to which emission in the soft X-ray and UV/optical bands lags that in the hard X-ray band. Fig.~\ref{lag_spec} then shows lags relative to the UVW2 band. Power-law and linear models describe this lag spectrum equally well statistically. The power-law fit follows the 4/3 rule ($\tau \propto ~\lambda^{1.36\pm13}$) expected for a standard thin disc, but a theoretical model for a standard thin disc that includes both viscous and X-ray heating predicts lags of smaller values than observed. Thus Fig.~\ref{lag_spec} suggests support for a more complex picture in which the accretion disc may be larger than in a standard thin-disc model.

\section{Results and Discussion}
We have studied UV/optical and X-ray variability of the bare Seyfert~1 galaxy Fairall~9 using intensive \swift{} monitoring  over four epochs lasting 2 to 6 months and spanning nearly two years during 2013 to 2015. We utilized various variability tools to study the observed lightcurves. We  studied linear correlation between the hard ($1.5-5\kev$) and one of the low energy bands -- soft ($0.5-1.5\kev$) and six UV/optical bands, and used the count-count correlation with positive offset method to investigate the slowly variable component. We have also studied cross-correlation using ZDCF and {\tt JAVELIN} and calculated the lag between various energy bands. Below we list the main results results followed by a summary which are as follows:  
\begin{enumerate}[(i)]

\item Fairall~9 exhibited strong variability in all bands. The hard ($1.5-5\kev$) and soft ($0.5-1.5\kev$) bands varied by factors of about four while the UV/optical bands UVW2 and B bands varied by $\sim 179\%$ and $\sim 73\%$, respectively, over the two year long monitoring period. 

\item Fairall~9 also showed short-term variability by a factor $\sim 2$ in the X-ray bands, and variations at a level of $\sim 60-7\%$ in the UV/optical bands on scales of days.

\item Pearson's $r$ coefficient for linear correlation between the hard band and lower energy bands are quite large ($0.63-0.92$). The significance of each correlation is very high which suggests that the observed lightcurves in the UV/optical to X-ray bands are strongly correlated. 

\item The cross-correlation analysis has revealed that the variations in the UV/optical bands are delayed with respect to the variations in the $1.5-5\kev$ X-ray band by $\sim 2-10{\rm~days}$. The lags measured with ZDCF and {\tt JAVELIN} for each pair of lightcurves are consistent within errors. 

\item The lag spectrum (lag vs. effective wavelength) shows that the lag increases with wavelength. The lag spectrum is described equally well by both the power-law model (see Equation~\ref{equ:lag_wv}) and a linear function. The simple models could not be distinguished possibly due to the large errors in the observed lags.

\item The predicted lags from standard disc theory with X-ray heating are shorter than the observed lags (see Fig.~\ref{lag_spec}).
\end{enumerate}

The optical/UV and X-ray lightcurves appear to show the long and short term variations by a factor of $\sim$four and $\sim$two, respectively. The variations are larger in the X-rays than that of the UV/optical emission. These variations on long and short timescales reflect the driving mechanism either related to intrinsic variations in the disc or heating of the accretion disc's surface externally by the X-ray source above the accretion disc. The variations appear strongly correlated. The lightcurves in the epoch II+III clearly show that the X-rays are leading  the UV/optical emission (see vertical lines in Fig.~\ref{iipiii}).

The cross-correlation analysis confirms that the variations in the UV/optical bands are delayed with respect to the variations in the X-ray band. This clearly suggests that variability associated with the changes in the accretion flow is not dominating. There are slowly rising long trends present in the observed X-ray/UV/optical lightcurves i.e., after 850 days of MJD-56000. The slow variations may be caused due to the changes intrinsically to the disc on viscous timescale. However, absence of clear optical/UV leads suggests that even the slow variations are not associated with intrinsic changes in the disc. The presence of UV/optical lags clearly implies that the X-ray heating dominates UV/optical emission and the variations in X-rays echo in each of the UV/optical bands.

Indeed, the X-ray heating in the accretion disc predicts lower lag than the observed lags. We used black hole mass $M_{BH}=2.6\times10^8~M_{\odot}$, height $H=6r_{g}$, inner radius $R_{in}=6r_{g}$, $\dot{m_{E}}=0.02$ and $L_{X}=10^{44.8}$\ergsec ~to estimate the theoretical lags which is shown by the green dashed line in Fig.~\ref{lag_spec}. These predicted lags, however, are overestimated as we did not account for the flux contributed from disc radii smaller than that implied by the Wien's law. Thus the wavelength obtained from Wein's law artificially provides longer lag from the accretion disc. Therefore, including geometrical effects such as flux weighted radius can further lower the predicted lags from the standard accretion disc model.  Such geometrical effects are included in the findings of \citet{2014MNRAS.444.1469M}. They found nearly three times longer observed lags than the predicted lag from standard accretion disc model. Thus the predicted lags will be shorter than depicted by the green line in Fig.~\ref{lag_spec}.  The discrepancy between the observed lags from Fairall~9 and the predicted lags from the standard disc model will be even more than shown in Fig.~\ref{lag_spec}. This implies that the emission regions in AGN discs are larger than that inferred from the standard accretion disc models. We note that micro-lensing studies e.g., \citep{2010ApJ...712.1129M} also require larger emission region for a given wavelength over the standard accretion disc. \citet{2011ApJ...727L..24D}  suggested that the inhomogeneities in accretion disc can provide the longer lag as observed because the outer inhomogeneous disc can emit higher flux responsible for larger region.

Despite the low predicted lags, the disc modeling provides important insights into the emitting regions of an accretion disc. If the accretion disc is face on, then the normalization expressed by Equation~\ref{equ:alpha} can give the radius of annulus emitting region corresponding to the reference wavelength $\lambda_{0}$ (e.g., \citealt{2015ApJ...806..129E}).  The radius of the accretion disc is found to be $r/R_{S}\sim400$ from the centre of the SMBH peaking at $1.5\times10^3~K$. Thus, the disc modeling provides important fact such as size of emitting region of the accretion disc.

Additionally further investigation of lag spectrum, the linear model can also describe the observed lags. This may be suggestive that the observed lags are affected by the viscous fluctuations in the accretion flow to give flatter distribution of lags. However, the timescales related to the accretion flow and X-ray heating to the disc are quite different. Both the linear and power-law models explain the lag spectrum equally well due to large error bars. Thus, given the large uncertainties in the measured lags, it is hard to distinguish between power-law and linear models. High signal-to-noise data can be helpful to address this issue in future. Essentially, our results indicate the complex nature of the UV/optical emitting regions of accretion disc, but have the potential to reveal important insights of central accretion disc and corona. A comprehensive observational monitoring from X-ray to UV/optical emission may provide much clearer picture of the disc regions. This campaign can be possible with the \astrosat{} mission \citep{2014SPIE.9144E..1SS} which has the capability to observe simultaneously across multiple bands from optical to hard X-rays.
 
\section{Acknowledgement}
   The authors thank the anonymous referee for his/her suggestions which improved the manuscript significantly. MP thanks the financial support of CSIR, New Delhi for this work. We thank the \swift{}~helpdesk and the help of Prof. Paul Kuin to understand the UVOT data reduction. We are also thankful to Dr. Pankaj Kushwaha for Python support. This research has made use of archival data of \swift{} observatory. We also used the software and/or web tools obtained from the High Energy Astrophysics Science Archive Research Center (HEASARC) of the NASA/GSFC Astrophysics Science Division and of Smithsonian Astrophysical Observatory's High Energy Astrophysics Divis
ion.

\newcommand{\pasp}{PASP} \def\apj{ApJ} \def\mnras{MNRAS}
\def\aap{A\&A} \def\apjl{ApJ} \def\aaps{A\&AS}\def\aj{aj} \def\physrep{PhR}
\def\pre{PhRvE} \def\apjs{ApJS} \def\pasa{PASA} \def\pasj{PASJ}
\def\nat{Nat} \def\ssr{SSRv} \def\aapr{AAPR} \def\araa{ARAA} \def\procspie{SPIE}

\bibliographystyle{mn2e} 
\bibliography{refs}

\end{document}